\documentclass[12pt]{article}
\usepackage{amsmath,amssymb,theorem,cite,epsfig,url,psfrag,eepic,amsmath,mathtools}
\advance \topmargin by -\headheight
\advance \topmargin by -\headsep
\evensidemargin \oddsidemargin
\def\Maketitle{{\def\newpage{}\maketitle}}
\makeatletter
\def\Appendix{\appendix
  \def\@seccntformat##1{Appendix~\csname the##1\endcsname.~~}}
\makeatother
\makeatletter
\@addtoreset{equation}{section}

\makeatother

\def\XXint#1#2#3{{\setbox0=\hbox{$#1{#2#3}{\int}$}
\vcenter{\hbox{$#2#3$}}\kern-.5\wd0}}

\begin{document}
\vspace*{0.3cm}
\title{\textbf{On dual description\\ of the deformed $O(N)$ sigma model}\vspace*{.3cm}}
\author{A.~V.~Litvinov$^{1,2}$ and L.~A.~Spodyneiko$^{1,3}$\\[\medskipamount]
\parbox[t]{0.85\textwidth}{\normalsize\it\centerline{1. Landau Institute for Theoretical Physics, 142432 Chernogolovka, Russia}}\\
\parbox[t]{0.85\textwidth}{\normalsize\it\centerline{2. National Research University Higher School of Economics,119048 Moscow, Russia}}\\
\parbox[t]{0.85\textwidth}{\normalsize\it\centerline{3. California Institute of Technology, Pasadena, CA 91125, USA}}
}
\date{}
\Maketitle
\abstract{We study dual strong coupling description of integrability-preserving deformation of the $O(N)$ sigma model. Dual theory is described by a coupled theory of Dirac fermions with four-fermion interaction and bosonic fields with exponential interactions. We claim that both  theories share the same integrable structure and coincide as quantum field theories. We construct a solution of Ricci flow equation which behaves in the UV as a free theory perturbed by graviton operators and show that it coincides with the metric of the $\eta-$deformed $O(N)$ sigma-model after  $T-$duality transformation.}
\section{Introduction}
Studying quantum field theories in the strong coupling regime is one of the most important problems of modern theoretical physics. In general, there are no methods for doing this, but a new tool arises when the quantum system admits the so called weak/strong coupling duality. In this case, the theory admits two perturbative expansions with coupling constants being inversely related. Typically, these descriptions use microscopic degrees of freedom and actions of completely different nature. One of the most notable examples of the duality, known as bosonization in two spacetime dimensions,  was pioneered by Coleman and Mandelstam \cite{Coleman:1974bu,Mandelstam:1975hb}.  One part of the duality is described by a massive Dirac fermion with four-fermion interaction, while the other by a scalar field with the cosine potential, the celebrated sine-Gordon model. Both theories were shown to coincide at the level of perturbation theory in the mass parameter. Other examples of dualities usually mentioned in the literature include electric-magnetic dualities in supersymmetric gauge theories, various string dualities, AdS/CFT duality etc.  

The main purpose of this paper is to motivate a weak/strong coupling duality between two integrable quantum field theories in two space-time dimensions. The former theory is the so called  $\eta-$deformed $O(N)$ non-linear sigma model \cite{Klimcik:2008eq,Delduc:2013fga}. The latter is effectively described by the coupled theory of Dirac fermions with four-fermion interaction (of Thirring type) and bosonic fields with exponential interaction (of Toda type). This theory has been recently introduced in \cite{Fateev:2018yos}.  Strong coupling regime of the sigma-model coincides with the perturbative regime of the Toda-Thirring model and vice versa. Below we formulate the results in more details.

We will treat the $O(N)$ sigma model via the coset construction, as a gauging of the two-dimensional principal chiral field model (PCF). The PCF is the non-linear sigma-model whose target space is a group manifold $G$.  The action of the PCF model is
\begin{equation}\label{PCF-action}
   \mathcal{S}=\frac{1}{2}\int\textrm{Tr}\left(\mathbf{g}^{-1}\partial_{+}\mathbf{g},\mathbf{g}^{-1}\partial_{-}\mathbf{g}\right) d^{2}x,
\end{equation}
where $\mathbf{g}$ is a map from the 2D space-time to the simple Lie group $G$ and $\partial_{\pm}$ are the light-cone derivatives. This theory has a global $G_{L}\times G_{R}$ symmetry which acts as
\begin{equation}
    \mathbf{g}\rightarrow U\mathbf{g}V,\qquad U,V\in G.
\end{equation}
Moreover, the theory \eqref{PCF-action} is known to be classically integrable \cite{Zakharov:1973pp}. The simplest way to show the integrability is to notice that the left Noether current $\mathbf{J}_{\pm}=\mathbf{g}\partial_{\pm}\mathbf{g}^{-1}$ is  both conserved and flat
\begin{equation}\label{JJ-2equation}
    \begin{aligned}
        &\partial_{+}\mathbf{J}_{-}+\partial_{-}\mathbf{J}_{+}=0,\\
        &\partial_{+}\mathbf{J}_{-}-\partial_{-}\mathbf{J}_{+}+[\mathbf{J}_{+},\mathbf{J}_{-}]=0.
    \end{aligned}
\end{equation}
These two equations can be rewritten as a compatibility condition known as a zero curvature representation\cite{Zakharov:1973pp}
\begin{equation}\label{ZCR}
   [\mathbf{D}_{+},\mathbf{D}_{-}]=0,\quad\text{where}\quad\mathbf{D}_{\pm}=\partial_{\pm}-\frac{\mathbf{J}_{\pm}}{1\pm\lambda},
\end{equation}
and $\lambda$ is an arbitrary complex number (spectral parameter). As a simple consequence of the zero curvature representation \eqref{ZCR}, the Wilson loop built out of this connection generates an infinite tower of conserved quantities, the integrals of motion. This infinite set of quantities can be used to solve the model by means of the inverse scattering method.  It has been argued in \cite{Polyakov:1983tt} and subsequently passed many consistency checks that the theory defined by the action \eqref{PCF-action} is integrable at the quantum level. Quantum systems possessing this property are usually handled within the framework of the Quantum Inverse Scattering Method \cite{Faddeev:1979gh}. However, this method is known to fail when applied to integrable non-linear sigma models directly. Only recently a considerable progress has been achieved in this direction (see \cite{Bazhanov:2017nzh} and discussions therein).

Now, let $H$ be the Lie subgroup of $G$ and $\mathfrak{h}$ be the corresponding Lie algebra, such that the quotient manifold is a symmetric space. We can define the sigma-model on this symmetric space by gauging the left symmetry of the action  \eqref{PCF-action}
\begin{equation}
    \partial_{\pm}\rightarrow D_{\pm}=\partial_{\pm}-A_{\pm},\quad A_{\pm}\in\mathfrak{h}.
\end{equation}
Clearly, this procedure breaks the left symmetry, but preserves the right one. It can be shown that the model is still enjoys the zero curvature representation and hence possesses the integrability property. In this paper we consider the case of $O(N)$ sigma-model, i.e. we take $G=SO(N)$ and $H=SO(N-1)$. Integrability of this sigma-model at the quantum level has been first demonstrated by Polyakov \cite{Polyakov:1977vm}. As QFT $O(N)$ sigma-model corresponds to an asymptotically free theory with a dynamically generated mass scale. It describes scattering of $N$ mesons in the vector representation of the global $O(N)$ group \cite{Polyakov:1975rr,Brezin:1975sq}. The scattering matrix for these mesons is strongly constrained by integrability, which implies, in particular,  the absence of particle production and factorization of the multi-particles amplitudes into the product of the two-particle scattering.  These requirements plus the conditions of crossing invariance and unitarity are so strong that allow one to compute the $S$-matrix exactly. The two-particle $S-$matrix for the $O(N)$ sigma-model has been found by Alexander and Alexey Zamolodchikov in their seminal paper \cite{Zamolodchikov:1978xm}. It has an explicit form
\begin{equation}\label{ZZ-Smatrix}
\begin{gathered}
     S_{ij}^{kl}(\theta)=\delta_{ij}\delta_{kl}S_{1}(\theta)+\delta_{ik}\delta_{jl}S_{2}(\theta)+\delta_{il}\delta_{jk}S_{3}(\theta),\\
   S_{3}(\theta)=S_{1}(i\pi-\theta),\quad S_{1}(\theta)=-\frac{2i\pi}{(N-2)(i\pi-\theta)}S_{2}(\theta),\quad S_{2}(\theta)=Q(\theta)Q(i\pi-\theta),
\end{gathered}
\end{equation}
where $Q(\theta)=\left(\Gamma\left(\frac{1}{N-2}-\frac{i\theta}{2\pi}\right)\Gamma\left(\frac{1}{2}-\frac{i\theta}{2\pi}\right)\right)/
   \left(\Gamma\left(\frac{1}{2}+\frac{1}{N-2}-\frac{i\theta}{2\pi}\right)\Gamma\left(-\frac{i\theta}{2\pi}\right)\right)$, and $\theta=\theta_{1}-\theta_{2}$ is the rapidity difference of the incoming states.

The $S-$matrix given above is known as the rational solution to the Yang-Baxter equation. It means that all matrix elements are rational functions of the rapidity (except for the prefactor). These solutions are known to be associated with the certain class of infinite dimensional algebras called the Yangians, in our case the Yangian of $O(N)$. It is known that the Yangians always admit one parametric deformation called the quantum affine group. The corresponding solution to the Yang-Baxter equation is the trigonometric solution, i.e. the matrix elements are expressed in terms of trigonometric functions.  For example, the trigonometric deformation of the $O(3)$ symmetric $S-$matrix has been found in \cite{Zamolodchikov:1980ku}. It describes the scattering of two charged and one neutral particles $(A_{\pm},A_{0})$. Explicitly, this $S-$matrix has the form (here $\lambda$ is the deformation parameter)
\begin{equation}\label{S-sausage}
\begin{gathered}
    S_{++}^{++}(\theta)=\frac{\sinh\lambda(\theta-i\pi)}{\sinh\lambda(\theta+i\pi)},\quad
   S_{+0}^{+0}=\frac{\sinh\lambda\theta}{\sinh\lambda(\theta-2i\pi)}S_{++}^{++}(\theta),\quad
    S_{00}^{00}(\theta)=S_{+0}^{+0}(\theta)+S_{-+}^{+-}(\theta),\\
   S_{-+}^{+-}(\theta)=-\frac{\sin\pi\lambda\sin2\pi\lambda}{\sinh\lambda(\theta-2i\pi)\sinh\lambda(\theta+i\pi)},\quad
   S_{+0}^{0+}=-\frac{i\sin2\pi\lambda}{\sinh\lambda(\theta-2i\pi)}S_{++}^{++}(\theta),
\end{gathered}
\end{equation}
with all other matrix elements related by $\mathbf{CPT}$-symmetry and crossing invariances. In the limit $\lambda\rightarrow0$ one recovers the rational $S-$matrix \eqref{ZZ-Smatrix} for $N=3$ in the basis $(\frac{(A_{+}+iA_{-})}{\sqrt{2}},\frac{(A_{+}-iA_{-})}{\sqrt{2}},A_{0})$. The generic $O(N)$ trigonomentric solution to the Yang Baxter equation has been constructed in \cite{Jimbo:1985ua,Bazhanov:1984gu}. It has the form very similar to \eqref{S-sausage}. It is important that this $S-$matrix similarly to \eqref{S-sausage} depends on a continuous parameter $\lambda$ such that in the limit $\lambda\rightarrow0$ one recovers Zamolodchikov's $S-$matrix of the $O(N)$ model \eqref{ZZ-Smatrix}.

It is tempting to find an appropriate deformation of the $O(N)$ sigma-model such that the scattering in the deformed theory is described by the trigonometric $O(N)$ $S-$matrix.   This deformation, if exists, must correspond to the renormalizable quantum field theory. The one-loop renormalizability  of the non-linear sigma model\footnote{In general, there might be other local terms in the action such as $B-$field or dilaton. We will be forced to include them later.}
\begin{equation}\label{SM-action-general}
    \mathcal{A}=\frac{1}{4\pi}\int G_{\mu\nu}(X)\partial X^{\mu}\bar{\partial}X^{\nu}d^{2}x,
\end{equation}
requires the target space metric $G_{\mu\nu}=G_{\mu\nu}(X)$ to satisfy the Ricci flow equations \cite{Friedan:1980jm}
\begin{equation}\label{Ricci-general}
    R_{\mu\nu}=-\dot{G}_{\mu\nu},
\end{equation}
where the derivative is taken with respect to the RG time $t$ the logarithm of the scale. The solutions of this non-linear evolution equation corresponding to the local quantum field theory must be UV stable, i.e. has to have a smooth limit as $t\rightarrow-\infty$. In general, if one starts with an arbitrarily chosen metric at some intermediate scale $t_{0}$, the solution will blow up before reaching the UV region.  The existence of UV stable solutions is a rather nontrivial fact and there are only a few explicitly known examples satisfying this property. Most of them correspond to Einstein manifolds such as round sphere $S_{N-1}$ in the case of $O(N)$ sigma-model. One of the exceptions is the one-parametric deformation of the round $2-$sphere known as the ``sausage'' metric \cite{Fateev:1992tk}
\begin{equation}\label{Sausage-metric}
   ds^{2}=\frac{\kappa}{\nu}\left(\frac{d\zeta^{2}}{(1-\zeta^{2})(1-\kappa^{2}\zeta^{2})}+\frac{(1-\zeta^{2})d\phi^{2}}{(1-\kappa^{2}\zeta^{2})}\right).
\end{equation}
Here $\phi$ is the $U(1)$ isometry coordinate $\phi\in[0,2\pi]$ and $\zeta\in[-1,1]$ is the longitudinal coordinate along the sausage. The parameter $\kappa$ is the running coupling constant
\begin{equation}
   \kappa=-\tanh \nu t,
\end{equation}
and $\nu$ is the deformation parameter. As $\nu$ goes to zero the metric \eqref{Sausage-metric} becomes that of the round $2-$sphere. For generic $\nu$ the metric  \eqref{Sausage-metric} can be visualized as an embedding of the sausage of length $L=-\sqrt{\nu}t$ and circumference $r=\sqrt{\frac{\kappa}{\nu}}$ into three-dimensional space-time. In the UV limit $t\rightarrow-\infty$ the sausage looks like an infinite cylinder corresponding to the asymptotically free theory.  Contrary, in the limit $t\rightarrow0$ the sausage behaves like a shrinking sphere, where the theory becomes strongly interacting and the perturbative expansion is no longer valid. It was conjectured and checked in \cite{Fateev:1992tk} that the non-linear sigma model defined by the one-loop metric \eqref{Sausage-metric} provides an integrable deformation of the  $O(3)$ sigma model with the  trigonometric $S-$matrix \eqref{S-sausage}. The deformation parameter $\lambda$ of the $S-$matrix is related to the parameter $\nu$ as
\begin{equation}
   \lambda=\nu+O(\nu^{2}).
\end{equation}
It is believed that the full theory including all higher loop corrections could be consistently extended starting from by the one-loop metric \eqref{Sausage-metric} in such a way that the resulting quantum field theory is integrable and described by the trigonometric $S-$matrix \eqref{S-sausage}.

Classical integrability of the sausage sigma model and its three-dimensional cousin has been demonstrated by Lukyanov \cite{Lukyanov:2012zt}. As it became clear recently, the sausage sigma model belongs to a more general class of the so called $\eta$-deformed sigma models.  It's all started with the seminal Klim\v c\'ik's  paper \cite{Klimcik:2008eq} where he suggested one-parametric deformation of the PCF action \eqref{PCF-action} preserving the integrability property. The deformation is done with the help of the linear skew-symmetric operator $\mathcal{R}$ on the complexified Lie algebra $\mathfrak{g}$ of the group $G$ which satisfies the modified Yang-Baxter relation
\begin{equation}\label{mYB-eq}
   [\mathcal{R}a,\mathcal{R}b]-\mathcal{R}([a,\mathcal{R}b]+[\mathcal{R}a,b])-[a,b]=0.
\end{equation}
With the operator $\mathcal{R}$ at hand the deformed action PCF is defined as
\begin{equation}\label{PCF-action-deformed}
   \mathcal{S}=\frac{1}{2}\int\textrm{Tr}\left(\mathbf{g}^{-1}\partial_{+}\mathbf{g}\,\frac{1}{1-\eta\mathcal{R}}\,\mathbf{g}^{-1}\partial_{-}\mathbf{g}\right) d^{2}x,
\end{equation}
where $\eta$  is the deformation parameter, at $\eta=0$ we return to the original action \eqref{PCF-action}. For $\eta\neq0$ the global symmetry group is broken down to $G_{L}\times N$, where $N$ is the subgroup of $G_{R}$ which commutes with the operator $\mathcal{R}$. In the virtue of the relation \eqref{mYB-eq}  the left Noether current
\begin{equation}
    \mathbf{J}_{\pm}^{\eta}=-\mathbf{g}\left(\frac{1}{1\pm\eta\mathcal{R}}\mathbf{g}^{-1}\partial_{\pm}\mathbf{g}\right)\mathbf{g}^{-1},
\end{equation}
is flat and hence the deformed theory admits the zero curvature representation leading to the classical integrability. 

We note, that since the left global symmetry $G_{L}$ is unaffected one can still apply the coset construction. As it was shown in \cite{Delduc:2013fga} the resulting theory is integrable provided the quotient space $G/H$ is a symmetric space. The action of the $\eta-$deformed coset sigma model can be written in the form \cite{Delduc:2013fga}
\begin{equation}\label{Coset-action-deformed}
   \mathcal{S}=\frac{1}{2}\int\textrm{Tr}\left(
   \left(\mathbf{g}\partial_{+}\mathbf{g}^{-1}\right)^{(\textrm{c})}\,\frac{1}{1-\eta\mathcal{R}_{\mathbf{g}}\circ\mathrm{P}_{\textrm{c}}}\,
   \left(\mathbf{g}\partial_{-}\mathbf{g}^{-1}\right)^{(\textrm{c})}\right) d^{2}x,
\end{equation}
where $\mathcal{R}_{\mathbf{g}}=\textrm{Ad}\, \mathbf{g}\circ\mathcal{R}\circ\textrm{Ad}\,\mathbf{g}^{-1}$ and $\mathrm{P}_{\textrm{c}}$ is the projection on the coset space.

So far we have not specified an explicit form of the operator $\mathcal{R}$ satisfying the bilinear relation \eqref{mYB-eq}. The common choice which has been widely elaborated in the literature since \cite{Klimcik:2008eq} requires  the standard decomposition of the Lie algebra $\mathfrak{g}$
\begin{equation}\label{g-decomposition}
   \mathfrak{g}=\mathfrak{c}\oplus_{\alpha>0}\mathfrak{g}_{\alpha}\oplus_{\alpha>0}\mathfrak{g}_{-\alpha},
\end{equation}
with the operator $\mathcal{R}$ defined by
\begin{equation}
  \mathcal{R}\Bigl|_{\mathfrak{c}}=0,\qquad
  \mathcal{R}\Bigl|_{\mathfrak{g}_{\alpha}}=i,\qquad
  \mathcal{R}\Bigl|_{\mathfrak{g}_{-\alpha}}=-i.
\end{equation}
Below, it will be convenient to work with the action \eqref{Coset-action-deformed} in an explicit coordinate system. We will use the standard basis for $SO(N)$ algebra
\begin{equation}
   (T_{ab})_{ij}\overset{\text{def}}{=}\delta_{ai}\delta_{bj}-\delta_{bi}\delta_{aj},
\end{equation}
and set   $T_{12},\; T_{34},\; T_{56},\dots$ to be the Cartan subalgebra. The $SO(N-1)$ subalgebra is spanned by the generators $T_{ab}$ with $a,b\neq1$. First non-trivial case corresponds to the $SO(3)/SO(2)$ coset. We use the following parametrization for the coset representative $\mathbf{g}$
\begin{equation}\label{SO(3)-parametrization}
    \mathbf{g}^{-1}=\exp\left(\phi\, T_{12}\right)\exp\left(\arcsin \zeta\,T_{13}\right).
\end{equation}
Then, one can show \cite{Hoare:2014pna} that the action  \eqref{Coset-action-deformed} with $\mathbf{g}$ given by \eqref{SO(3)-parametrization} leads to the sigma model with the metric \eqref{Sausage-metric} with $\eta=i\kappa$. This suggests that the theory \eqref{Coset-action-deformed} for general cosets can be treated as a natural generalization of the sausage sigma-model and that the corresponding $S$-matrix coincides with the trigonometric deformation of the $O(N)$ sigma-model.

Now, we come to another important point. The trigonometric $S-$matrix  \eqref{S-sausage} admits a perturbative expansion around the point $\lambda=\frac{1}{2}$. Exactly at the value $\lambda=\frac{1}{2}$ the model reduces to three non-interacting particles $(A_{0},A_{+},A_{-})$, which can be effectively described by  free scalar $\Phi$ and  Dirac fermion $\psi$ fields of the same mass. For the interacting theory, as can be shown order by order, the perturbative expansion of the $S-$matrix \eqref{S-sausage} coincides with the one obtained from the Langrangian\footnote{It was first noticed by Alyosha Zamolodchikov (unpublished).}
\begin{equation}\label{Lagrangian-Sausage}
    \mathcal{L}=\frac{1}{8\pi}\bigl(\partial_{\mu}\Phi\bigr)^{2}+i\bar{\psi}\gamma^{\mu}\partial_{\mu}\psi+
    \frac{\pi b^{2}}{2(1+b^{2})}\bigl(\bar{\psi}\gamma^{\mu}\psi\bigr)^{2}-m\bar{\psi}\psi\cosh(b\Phi)-\frac{m^{2}}{8\pi b^{2}}\sinh^{2}(b\Phi),
\end{equation}
provided that the parameter $\lambda$ from the $S-$matrix \eqref{S-sausage}  and the parameter $b$ from the Lagrangian \eqref{Lagrangian-Sausage} are related according to
$\lambda=\frac{1}{2(1+b^{2})}$. The existence of the dual description of the sausage sigma model is a rather miraculous phenomenon. In this paper, we partially shed light on the origin of this relation. The short explanation: this is the integrability who stands behind the duality. More precisely, in some limit both theories can be shown to share the same integrable structure and the integrability property is so strong that the theories stay equivalent even beyond the limit. Here we should stress that the Lagrangian \eqref{Lagrangian-Sausage} is exact in the parameter $b$, meaning that in the perturbative region $b\rightarrow0$ or $\lambda\rightarrow\frac{1}{2}$ it describes the full theory. In the strong coupling regime $b\rightarrow\infty$ the Lagrangian \eqref{Lagrangian-Sausage} is certainly useless, but instead, we have a sigma model description with the metric \eqref{Sausage-metric}, which, however, is accurate only up to one-loop order.

The outline of the paper is the following. In section \ref{dual-L} we introduce following \cite{Fateev:2018yos} the generalization of the Lagrangian \eqref{Lagrangian-Sausage} depending on a continuous parameter $b$ for arbitrary $N$. We treat this theory as a perturbed conformal field theory and find a part of an infinite conformal symmetry, a system of local Integrals of Motion, which survives the perturbation and has the $O(N-1)$ symmetry in the limit $b\rightarrow\infty$. We study corresponding conformal field theory in section \ref{CFT}.  In section \ref{ricci}, we study very peculiar property of the system under consideration which is the generalization of the $b\rightarrow\frac{1}{b}$ symmetry of Toda theories. Namely, we use the fact that the same system of Integrals of Motion can be obtained from the perturbed CFT of sigma model type. This theory, however, is not UV finite and requires renomalization. Treating it in the vicinity of the trivial fixed point, we study the Ricci flow equations \eqref{Ricci-general} with the prescribed UV asymptotic and find an exact solution for the metric. We show that this metric coincides with the metric for the $\eta-$deformed coset sigma-model \eqref{Coset-action-deformed} after a coordinate change and $T-$duality transformations. In appendices we collect important facts and present details of some computations.
\section{Dual Lagrangians}\label{dual-L}
Let us start with known results and consider the theory defined by the Lagrangian density \eqref{Lagrangian-Sausage}. The last term
\begin{equation}
  -\frac{m^{2}}{8\pi b^{2}}\sinh^{2}(b\Phi),
\end{equation}
is of order $m^{2}$ and plays the role of the counter-term.  It can be shown that with this choice the theory does not have any divergencies and becomes UV finite. One may use Coleman-Mandelstam boson-fermion duality \cite{Coleman:1974bu,Mandelstam:1975hb}
\begin{equation}
    i\bar{\psi}\gamma^{\mu}\partial_{\mu}\psi\rightarrow\frac{1}{8\pi}(\partial_{\mu}\varphi)^{2},\qquad
    \bar{\psi}(1\pm\gamma_{5})\psi\rightarrow e^{\pm i\beta\varphi},\quad\text{where}\quad\beta=\sqrt{1+b^{2}},
\end{equation}
and rewrite the theory in the form
\begin{equation}\label{Lagrangian-Sausage-bosonic}
    \mathcal{L}=\frac{1}{8\pi}\bigl(\partial_{\mu}\varphi\bigr)^{2}+\frac{1}{8\pi}\bigl(\partial_{\mu}\Phi\bigr)^{2}-m\cos(\beta\phi)\cosh(b\Phi).
\end{equation}
The theory \eqref{Lagrangian-Sausage-bosonic} is a special integrable perturbation of Sine-Liouville conformal field theory \cite{Fateev:2017mug}. Namely, it means that it possesses infinitely many local integrals of motion of odd spins
\begin{equation}
   \mathbf{I}_{2k-1},\qquad \bar{\mathbf{I}}_{2k-1},
\end{equation}
which form a commutative set $[ \mathbf{I}_{2k-1}, \mathbf{I}_{2l-1}]=[ \mathbf{I}_{2k-1}, \bar{\mathbf{I}}_{2l-1}]=[ \bar{\mathbf{I}}_{2k-1}, \bar{\mathbf{I}}_{2l-1}]=0$.  One can check this statement perturbatively in the mass parameter $m$
\begin{equation}
   \mathbf{I}_{2k-1}=\mathbf{I}_{2k-1}^{\textrm{free}}+O(m),\quad\text{with}\quad
   \mathbf{I}_{2k-1}^{\textrm{free}}=\frac{1}{2\pi}\int_{\mathcal{C}}G_{2k}(\partial\varphi,\partial\Phi)dz,
\end{equation}
where $G_{2k}(\partial\varphi,\partial\Phi)$ is a holomorphic differential polynomial of degree $2k$. For example,
\begin{equation}
   G_{2}(\partial\varphi,\partial\Phi)=(\partial\varphi)^{2}+(\partial\Phi)^{2}.
\end{equation}
Higher-spin densities can be defined from the requirement  that they commute with the perturbation
\begin{equation}
   [\mathbf{I}_{2k-1}^{\textrm{free}},\int e^{\pm i\beta\varphi\pm b\Phi}dz]=0.
\end{equation}
Clearly, this equation is satisfied by $\mathbf{I}_{2}^{\textrm{free}}$, but its validity for $k>1$ is highly non-trivial. The integrability of the model is closely tied to the special combination of the fields in the exponents.

Motivated by the explicit form of the Lagrangian \eqref{Lagrangian-Sausage-bosonic} we will study more general setting. Let $\varphi=(\varphi_{1},\dots,\varphi_{N-1})$ be the $N-1$ component bosonic field and consider the theory defined by the Lagrangian density
\begin{equation}\label{Lagrangian}
  \mathcal{L}=\frac{1}{8\pi}\bigl(\partial_{\mu}\varphi\bigr)^{2}+\Lambda\sum_{r=1}^{N}e^{(\boldsymbol{\alpha}_{r},\varphi)},
\end{equation}
where $(\boldsymbol{\alpha}_{1},\dots,\boldsymbol{\alpha}_{N})$ is a given set of vectors\footnote{In general the number of vectors $\boldsymbol{\alpha}_{r}$ is bigger by one than the number of fields. There are exceptions from this rule at lower dimensions, such as the case $N=3$ considered above.}, which is required to have maximal rank. We are only interested in the quantum field theories of this form with infinitely many integrals of motion. In the leading order in $\Lambda$ this constrains the integrals of motion $\mathbf{I}_{k}^{\textrm{free}}$ to obey
\begin{equation}\label{free-IOM}
   [\mathbf{I}_{k}^{\textrm{free}},\int e^{(\boldsymbol{\alpha}_{r},\varphi)}dz]=0,\quad\text{for all}\quad r=1,\dots,N.
\end{equation}
These equalities (see appendix \ref{IM}) can be treated as both, the equation for unknown integrals of motion $\mathbf{I}_{k}^{\textrm{free}}$ and for the set of vectors $(\boldsymbol{\alpha}_{1},\dots,\boldsymbol{\alpha}_{N})$.  In fact, they are so strong, that the solution exits only if the Gram matrix $\Gamma_{r,s}=(\boldsymbol{\alpha}_{r},\boldsymbol{\alpha}_{s})$ takes a very special form. We will be interested in the case where all roots $\boldsymbol{\alpha}_{r}$ are fermionic, meaning that
\begin{equation}
  (\boldsymbol{\alpha}_{r},\boldsymbol{\alpha}_{r})=-1,\quad\text{for all}\quad r.
\end{equation}
Moreover, we assume that only odd-spin integrals of motion are present: $\mathbf{I}_{2k}^{\textrm{free}}=0$,  because this is what we expect for the theory which has $O(N)$ symmetry in some limit. With these conditions specified one can find a series of solutions for every $N\geq3$ which depends on a continuous parameter $b$ (see appendix \ref{IM} and \cite{Litvinov:2016mgi} for more details). Since the cases $N=2n+1$ and $N=2n+2$ are different we consider them separately.  Below, we restrict ourself to $N>4$. Two cases $N=3$ and $N=4$ are special and we comment on them in the end of the section. 

The Gram matrix (degenerate) of the vectors $(\boldsymbol{\alpha}_{1},\dots,\boldsymbol{\alpha}_{N})$ has to be of a very special form. It will be convenient to introduce Dynkin-like picture in order to represent it. We will use the following conventions
\begin{equation*}
\hspace*{-4cm}
\begin{picture}(200,10)(215,123)
    \Thicklines
    \unitlength 5pt
    \put(69,25){\circle{2}}
    \put(79,25){\circle{2}}
    \put(68.4,24,4){\line(1,1){1.2}}
    \put(68.4,25,6){\line(1,-1){1.2}}
    \put(78.4,24,4){\line(1,1){1.2}}
    \put(78.4,25,6){\line(1,-1){1.2}}
    \put(70,25){\line(1,0){8}}
    \put(73,26){$x$}
\end{picture}
\text{corresponds to the $2\times2$ block matrix}\quad
\begin{pmatrix}
   -1&x\\
   x&-1
\end{pmatrix},
\end{equation*}
while the crossed circles not connected by an edge correspond to orthogonal vectors.
\paragraph{The case $N=2n+1$:}  the root system corresponds to the diagram
\begin{equation}\label{Bn}
\begin{picture}(300,60)(260,110)
    \Thicklines
    \unitlength 5pt
    \put(48,32){\circle{2}}
    \put(48,18){\circle{2}}
    \put(54.4,24,4){\line(-1,-1){7}}
    \put(54.4,25,6){\line(-1,1){7}}
    \put(47.4,31.4){\line(1,1){1.2}}
    \put(47.4,18.6){\line(1,-1){1.2}}
    \put(48,19){\line(0,1){12}}
    \put(55,25){\circle{2}}
    \put(54.4,24,4){\line(1,1){1.2}}
    \put(54.4,25,6){\line(1,-1){1.2}}
    \put(66,25){\line(1,0){8}}
    \put(56,25){\line(1,0){8}}
    \put(65,25){\circle{2}}
    \put(64.4,24,4){\line(1,1){1.2}}
    \put(64.4,25,6){\line(1,-1){1.2}}
    \put(75,25){\circle{2}}
    \put(74.4,24,4){\line(1,1){1.2}}
    \put(74.4,25,6){\line(1,-1){1.2}}
    \put(76,25){\line(1,0){2}}
    \put(80,25){\circle{.2}}
    \put(81,25){\circle{.2}}
    \put(82,25){\circle{.2}}
    \put(83,25){\circle{.2}}
    \put(84,25){\circle{.2}}
    \put(85,25){\circle{.2}}
    \put(87,25){\line(1,0){2}}
    \put(90,25){\circle{2}}
    \put(89.4,24,4){\line(1,1){1.2}}
    \put(89.4,25,6){\line(1,-1){1.2}}
    \put(91,25){\line(1,0){8}}
    \put(100,25){\circle{2}}
    \put(99.4,24,4){\line(1,1){1.2}}
    \put(99.4,25,6){\line(1,-1){1.2}}
    \put(101,25){\line(1,0){8}}
    \put(110,25){\circle{2}}
    \put(109.4,24,4){\line(1,1){1.2}}
    \put(109.4,25,6){\line(1,-1){1.2}}
    \put(109.4,24,4){\line(1,1){8.2}}
    \put(109.4,25,6){\line(1,-1){8.2}}
    \put(117,32){\circle{2}}
    \put(117,18){\circle{2}}
    \put(116.4,32.6){\line(1,-1){1.2}}
    \put(116.4,17.4){\line(1,1){1.2}}
    \put(117,19){\line(0,1){12}}
    \put(55,23){$\underbrace{\phantom{aaaaaaaaaaaaaaaaaaaaaaaaaaaaaaaaaaaaaaaaaaaaaaaa}}$}
    \put(80.5,18){$\scriptstyle{2n-3}$}
    \put(50.5,19){$\scriptstyle{-b^{2}}$}
    \put(50.5,30){$\scriptstyle{-b^{2}}$}
    \put(42,24.5){$\scriptstyle{1+2b^{2}}$}
    \put(57.5,26){$\scriptstyle{1+b^{2}}$}
    \put(68,26){$\scriptstyle{-b^{2}}$}
    \put(103,26){$\scriptstyle{-b^{2}}$}
    \put(92.5,26){$\scriptstyle{1+b^{2}}$}
    \put(110,19){$\scriptstyle{1+b^{2}}$}
    \put(110,30){$\scriptstyle{1+b^{2}}$}
    \put(117.5,24.5){$\scriptstyle{-1-2b^{2}}$}
    \put(45.5,16){$\scriptstyle{\boldsymbol{\alpha}_{1}}$}
    \put(45.5,33.5){$\scriptstyle{\boldsymbol{\alpha}_{2}}$}
     \put(54.6,27){$\scriptstyle{\boldsymbol{\alpha}_{3}}$}
     \put(64.6,27){$\scriptstyle{\boldsymbol{\alpha}_{4}}$}
     \put(74.6,27){$\scriptstyle{\boldsymbol{\alpha}_{5}}$}
    \put(118,16){$\scriptstyle{\boldsymbol{\alpha}_{2n}}$}
    \put(118,33.5){$\scriptstyle{\boldsymbol{\alpha}_{2n+1}}$}
  \end{picture}
  \vspace*{1cm}
\end{equation}
It is convenient to introduce  Cartesian coordinates in $\mathbb{R}^{2n}$: $(E_{i},e_{i})$, $i=1,\dots,n$ such that
\begin{equation}
   (E_{i},E_{j})=(e_{i},e_{j})=\delta_{ij},\quad (E_{i},e_{j})=0,
\end{equation}
and the vectors $\boldsymbol{\alpha}_{r}$ have the form
\begin{equation}\label{vectors}
 \begin{aligned}
   &  \boldsymbol{\alpha}_{1}=bE_{1}+i\beta e_{1},\qquad \boldsymbol{\alpha}_{2}=bE_{1}-i\beta e_{1},\\
   &   \boldsymbol{\alpha}_{2k-1}=-bE_{k-1}+i\beta e_{k},\qquad \boldsymbol{\alpha}_{2k}=bE_{k}-i\beta e_{k},\quad\text{for}\quad k=2,\dots,n,\\
   &   \boldsymbol{\alpha}_{2n}=bE_{n}-i\beta e_{n},\qquad \boldsymbol{\alpha}_{2n+1}=-bE_{n}-i\beta e_{n},
 \end{aligned}
\end{equation}
where  $\beta=\sqrt{1+b^{2}}$. It follows from \eqref{vectors} that
\begin{equation}
    \boldsymbol{\alpha}_{1}+ \boldsymbol{\alpha}_{2}+2\sum_{r=3}^{2n-1} \boldsymbol{\alpha}_{r}+\boldsymbol{\alpha}_{2n}+\boldsymbol{\alpha}_{2n+1}=0.
\end{equation}
In accordance with this representation it is convenient to define $\Phi_{k}=(\varphi,E_{k})$ and  $\phi_{k}=(\varphi,e_{k})$. We will study the theory in the region where $b\rightarrow0$. One has to add counter-terms in order to regularize the UV behavior\footnote{In the middle region $-1<b^{2}<0$ one can use an analytical regularization instead.}. Using the representation \eqref{Lagrangian} and the boson-fermion correspondence \cite{Coleman:1974bu,Mandelstam:1975hb} the Lagrangian \eqref{Lagrangian} can be rewritten as \cite{Fateev:2018yos}
\begin{multline}\label{Lagrangian2}
    \mathcal{L}=\sum_{k=1}^{n}\left(\frac{1}{8\pi}\bigl(\partial_{\mu}\Phi_{k}\bigr)^{2}+i\bar{\psi}_{k}\gamma^{\mu}\partial_{\mu}\psi_{k}+
    \frac{\pi b^{2}}{2(1+b^{2})}\bigl(\bar{\psi}_{k}\gamma^{\mu}\psi_{k}\bigr)^{2}\right)-\\-
    m\Bigl(e^{b\Phi_{1}}\bar{\psi}_{1}\psi_{1}+
    \sum_{k=2}^{n-1}\Bigl(e^{b\Phi_{k}}\bar{\psi}_{k}\Bigl(\frac{1+\gamma_{5}}{2}\Bigr)\psi_{k}+e^{-b\Phi_{k-1}}\bar{\psi}_{k}\Bigl(\frac{1-\gamma_{5}}{2}\Bigr)\psi_{k}\Bigr)
    +e^{-b\Phi_{n-1}}\bar{\psi}_{n}\Bigl(\frac{1-\gamma_{5}}{2}\Bigr)\psi_{n}+\\+\left.
    \cosh b\Phi_{n}\,\bar{\psi}_{n}\,\,\Bigl(\frac{1+\gamma_{5}}{2}\Bigr)\psi_{n}\right)-
    \frac{m^{2}}{8\pi b^{2}}\left(e^{2b\Phi_{1}}+2\sum_{k=2}^{n-1}e^{b(\Phi_{k}-\Phi_{k-1})}+e^{b(\Phi_{n}-\Phi_{n-1})}+e^{-b(\Phi_{n-1}+\Phi_{n})}\right),
\end{multline}
where $m=\Lambda/4\pi$.
The last term in \eqref{Lagrangian2} plays the role of the counter-term. We note that this part of the theory coincides with the  affine $B_{n}^{\vee}$ Toda field theory (the Dynkin diagram has the orientation of the arrows opposite to  $B_{n}$). One immediately sees that there are $n$ pairs of charged particles $\psi_{k}$, $\bar{\psi}_{k}$ and one scalar particle $\Phi_{n}$ of the same mass $M=m+O(b^{2})$. In order to describe the rest of the spectrum one has to diagonalize the $(n-1)\times(n-1)$ matrix
\begin{equation}
   A=\begin{pmatrix}
      3&-1&0&0&\hdotsfor{2}\\
      -1&2&-1&0&\hdotsfor{2}\\
      0&-1&2&-1&\hdotsfor{2}\\
      0&0&-1&2&\ddots&\dots\\
      \hdotsfor{3}&\ddots&\ddots&-1\\
      \hdotsfor{4}&-1&2
   \end{pmatrix}
\end{equation}
It is easy to check that
\begin{equation}
   \det\left(A-4\sin^{2} x\right)=\frac{\sin(2n-1)x}{\sin x},
\end{equation}
which implies that the masses of the remaining $(n-1)$ particles  have the form
\begin{equation}
    M_{k}=2m\sin\left(\frac{\pi k}{2n-1}\right)+O(b^{2}),\qquad k=1,2,\dots,n-1.
\end{equation}
\paragraph{The case $N=2n+2$:} the root system corresponds to the diagram
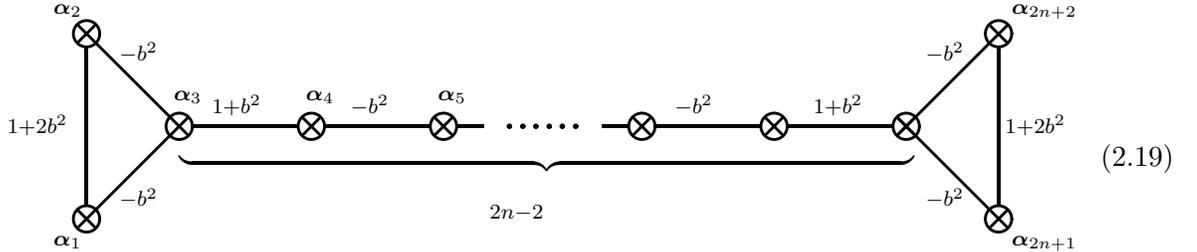
\begin{equation}\label{Dn}
\begin{picture}(300,60)(260,110)
    \Thicklines
    \unitlength 5pt
    \put(48,32){\circle{2}}
    \put(48,18){\circle{2}}
    \put(54.4,24,4){\line(-1,-1){7}}
    \put(54.4,25,6){\line(-1,1){7}}
    \put(47.4,31.4){\line(1,1){1.2}}
    \put(47.4,18.6){\line(1,-1){1.2}}
    \put(48,19){\line(0,1){12}}
    \put(55,25){\circle{2}}
    \put(54.4,24,4){\line(1,1){1.2}}
    \put(54.4,25,6){\line(1,-1){1.2}}
    \put(66,25){\line(1,0){8}}
    \put(56,25){\line(1,0){8}}
    \put(65,25){\circle{2}}
    \put(64.4,24,4){\line(1,1){1.2}}
    \put(64.4,25,6){\line(1,-1){1.2}}
    \put(75,25){\circle{2}}
    \put(74.4,24,4){\line(1,1){1.2}}
    \put(74.4,25,6){\line(1,-1){1.2}}
    \put(76,25){\line(1,0){2}}
    \put(80,25){\circle{.2}}
    \put(81,25){\circle{.2}}
    \put(82,25){\circle{.2}}
    \put(83,25){\circle{.2}}
    \put(84,25){\circle{.2}}
    \put(85,25){\circle{.2}}
    \put(87,25){\line(1,0){2}}
    \put(90,25){\circle{2}}
    \put(89.4,24,4){\line(1,1){1.2}}
    \put(89.4,25,6){\line(1,-1){1.2}}
    \put(91,25){\line(1,0){8}}
    \put(100,25){\circle{2}}
    \put(99.4,24,4){\line(1,1){1.2}}
    \put(99.4,25,6){\line(1,-1){1.2}}
    \put(101,25){\line(1,0){8}}
    \put(110,25){\circle{2}}
    \put(109.4,24,4){\line(1,1){1.2}}
    \put(109.4,25,6){\line(1,-1){1.2}}
    \put(109.4,24,4){\line(1,1){8.2}}
    \put(109.4,25,6){\line(1,-1){8.2}}
    \put(117,32){\circle{2}}
    \put(117,18){\circle{2}}
    \put(116.4,32.6){\line(1,-1){1.2}}
    \put(116.4,17.4){\line(1,1){1.2}}
    \put(117,19){\line(0,1){12}}
    \put(55,23){$\underbrace{\phantom{aaaaaaaaaaaaaaaaaaaaaaaaaaaaaaaaaaaaaaaaaaaaaaaa}}$}
    \put(78.5,18){$\scriptstyle{2n-2}$}
    \put(50.5,19){$\scriptstyle{-b^{2}}$}
    \put(50.5,30){$\scriptstyle{-b^{2}}$}
    \put(42,24.5){$\scriptstyle{1+2b^{2}}$}
    \put(57.5,26){$\scriptstyle{1+b^{2}}$}
    \put(68,26){$\scriptstyle{-b^{2}}$}
    \put(103,26){$\scriptstyle{1+b^{2}}$}
    \put(92.5,26){$\scriptstyle{-b^{2}}$}
    \put(111.5,19){$\scriptstyle{-b^{2}}$}
    \put(111.5,30){$\scriptstyle{-b^{2}}$}
    \put(117.5,24.5){$\scriptstyle{1+2b^{2}}$}
    \put(45.5,16){$\scriptstyle{\boldsymbol{\alpha}_{1}}$}
    \put(45.5,33.5){$\scriptstyle{\boldsymbol{\alpha}_{2}}$}
     \put(54.6,27){$\scriptstyle{\boldsymbol{\alpha}_{3}}$}
     \put(64.6,27){$\scriptstyle{\boldsymbol{\alpha}_{4}}$}
     \put(74.6,27){$\scriptstyle{\boldsymbol{\alpha}_{5}}$}
    \put(118,16){$\scriptstyle{\boldsymbol{\alpha}_{2n+1}}$}
    \put(118,33.5){$\scriptstyle{\boldsymbol{\alpha}_{2n+2}}$}
  \end{picture}
  \vspace*{1cm}
\end{equation}
In this case we  introduce the Cartesian coordinates in $\mathbb{R}^{2n+1}$: $E_{i}$, $i=1,\dots,n$ and $e_{j}$, $j=1,\dots,n+1$ such that
\begin{equation}
   (E_{i},E_{j})=(e_{i},e_{j})=\delta_{ij},\quad (E_{i},e_{j})=0,
\end{equation}
and the vectors $\boldsymbol{\alpha}_{r}$ have the form
\begin{equation}\label{vectors-2}
 \begin{aligned}
   &  \boldsymbol{\alpha}_{1}=bE_{1}+i\beta e_{1},\qquad \boldsymbol{\alpha}_{2}=bE_{1}-i\beta e_{1},\\
   &   \boldsymbol{\alpha}_{2k-1}=-bE_{k-1}+i\beta e_{k},\qquad \boldsymbol{\alpha}_{2k}=bE_{k}-i\beta e_{k},\quad\text{for}\quad k=2,\dots,n,\\
   &   \boldsymbol{\alpha}_{2n+1}=-bE_{n}+i\beta e_{n+1},\qquad \boldsymbol{\alpha}_{2n+2}=-bE_{n}-i\beta e_{n+1}.
 \end{aligned}
\end{equation}
Using the  boson-fermion correspondence \cite{Coleman:1974bu,Mandelstam:1975hb} the Lagrangian \eqref{Lagrangian}  for the $O(2n+2)$ model with $n\geq1$ can be rewritten as \cite{Fateev:2018yos}
\begin{multline}\label{Lagrangian3}
    \mathcal{L}=\sum_{k=1}^{n}\frac{1}{8\pi}\bigl(\partial_{\mu}\Phi_{k}\bigr)^{2}+\sum_{k=1}^{n+1}\left(i\bar{\psi}_{k}\gamma^{\mu}\partial_{\mu}\psi_{k}+
    \frac{\pi b^{2}}{2(1+b^{2})}\bigl(\bar{\psi}_{k}\gamma^{\mu}\psi_{k}\bigr)^{2}\right)-\\-
    m\Bigl(e^{b\Phi_{1}}\bar{\psi}_{1}\psi_{1}+
    \sum_{k=2}^{n}\Bigl(e^{b\Phi_{k}}\bar{\psi}_{k}\Bigl(\frac{1+\gamma_{5}}{2}\Bigr)\psi_{k}+e^{-b\Phi_{k-1}}\bar{\psi}_{k}\Bigl(\frac{1-\gamma_{5}}{2}\Bigr)\psi_{k}\Bigr)
    +e^{-b\Phi_{n}}\bar{\psi}_{n}\psi_{n}\Bigr)-\\-
    \frac{m^{2}}{8\pi b^{2}}\left(e^{2b\Phi_{1}}+2\sum_{k=2}^{n}e^{b(\Phi_{k}-\Phi_{k-1})}+e^{-2b\Phi_{n-1}}\right).
\end{multline}
We note that the purely bosonic  part of the theory coincides with the  affine $C_{n}$ Toda field theory. The spectrum consists of $2n+2$ charged particles $\bar{\psi}_{k}$, $\psi_{k}$ of the mass $M=m+O(b^{2})$. The spectrum of Toda part of the theory consists of $n$ particles with
\begin{equation}
   m_{k}=2m\sin\left(\frac{\pi k}{n}\right)+O(b^{2}).
\end{equation}

We conclude this section with a brief remark concerning $O(3)$ and $O(4)$ deformed sigma models and their dual descriptions. For $O(3)$ case the dual Lagrangian is given by \eqref{Lagrangian-Sausage} which has, after bosonization, four exponents in the perturbation, not $3$ as expected. This is a peculiar property of $N=3$ which does not happen for $N>3$. The case $N=4$ is also special. Because of an exceptional isomorphism $S^{3}\simeq SU(2)$, the $O(4)$ model can be treated equally as a coset sigma model or as a PCF model as well. According to Klim\v c\'ik \cite{Klimcik:2014bta} PCF admits two-parametric integrable deformation
\begin{equation}\label{PCF-action-double-deformed}
   \mathcal{S}=\frac{1}{2}\int\textrm{Tr}\left(\mathbf{g}^{-1}\partial_{+}\mathbf{g}\,\frac{1}{1-\alpha\mathcal{R}-\beta\mathcal{R}_{\mathbf{g}}}\,\mathbf{g}^{-1}\partial_{-}\mathbf{g}\right) d^{2}x,
\end{equation}
so that in particular  $O(4)$ sigma model can be doubly deformed while preserving integrability. This deformation and its dual description has been found in \cite{Fateev:1996ea}\footnote{The equivalence of Fateev's two-parameter deformation of the $O(4)$ SM with the doubly deformed Klim\v c\'ik sigma model has been demonstrated in \cite{Hoare:2014pna}.}. In these notes we consider $O(4)$ model as a coset sigma model, which in fact coincides with equal parameter deformation $\alpha=\beta$ of the PCF.
\section{Conformal field theory, reflection operator}\label{CFT}
The theory \eqref{Lagrangian} can be viewed as a perturbation of the conformal field theory obtained from \eqref{Lagrangian} by throwing away the last term $e^{(\boldsymbol{\alpha}_{N},\varphi)}$ from the Lagrangian. It is convenient to couple the conformal field theory to the background metric. The action in the curved space-time has the form
\begin{equation}\label{action}
   \mathcal{A}=\int\left(\frac{1}{8\pi}g^{ab}\left(\partial_{a}\varphi,\partial_{b}\varphi\right)+\frac{(\boldsymbol{\rho},\varphi)}{4\pi}R+
   \Lambda\sum_{r=1}^{N-1}e^{(\boldsymbol{\alpha}_{r},\varphi)}\right)\,\sqrt{g}\,d^{2}z,
\end{equation}
where $g_{ab}$ is the metric on a surface and $R$ is its scalar curvature. The vectors $\boldsymbol{\alpha}_{r}$ in \eqref{action} are exactly the same as in \eqref{vectors} and \eqref{vectors-2}. Clearly, they form a basis. The Gram matrix of these vectors depends on a continuous parameter $b$ and according to our conventions corresponds to the graph
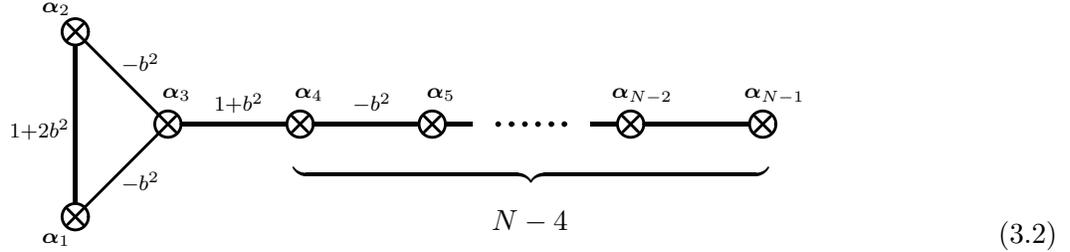
\begin{equation}\label{Dn-2}
\begin{picture}(300,100)(220,80)
    \Thicklines
    \unitlength 5pt
    \put(48,32){\circle{2}}
    \put(48,18){\circle{2}}
    \put(54.4,24,4){\line(-1,-1){7}}
    \put(54.4,25,6){\line(-1,1){7}}
    \put(47.4,31.4){\line(1,1){1.2}}
    \put(47.4,18.6){\line(1,-1){1.2}}
    \put(48,19){\line(0,1){12}}
    \put(55,25){\circle{2}}
    \put(54.4,24,4){\line(1,1){1.2}}
    \put(54.4,25,6){\line(1,-1){1.2}}
    \put(66,25){\line(1,0){8}}
    \put(56,25){\line(1,0){8}}
    \put(65,25){\circle{2}}
    \put(64.4,24,4){\line(1,1){1.2}}
    \put(64.4,25,6){\line(1,-1){1.2}}
    \put(75,25){\circle{2}}
    \put(74.4,24,4){\line(1,1){1.2}}
    \put(74.4,25,6){\line(1,-1){1.2}}
    \put(76,25){\line(1,0){2}}
    \put(80,25){\circle{.2}}
    \put(81,25){\circle{.2}}
    \put(82,25){\circle{.2}}
    \put(83,25){\circle{.2}}
    \put(84,25){\circle{.2}}
    \put(85,25){\circle{.2}}
    \put(87,25){\line(1,0){2}}
    \put(90,25){\circle{2}}
    \put(89.4,24,4){\line(1,1){1.2}}
    \put(89.4,25,6){\line(1,-1){1.2}}
    \put(91,25){\line(1,0){8}}
    \put(100,25){\circle{2}}
    \put(99.4,24,4){\line(1,1){1.2}}
    \put(99.4,25,6){\line(1,-1){1.2}}
    \put(64.5,22){$\underbrace{\phantom{aaaaaaaaaaaaaaaaaaaaaaaaaaaaaaa}}$}
    \put(79.5,17){$N-4$}
    \put(51.5,20){$\scriptstyle{-b^{2}}$}
    \put(51.5,29){$\scriptstyle{-b^{2}}$}
    \put(43,24){$\scriptstyle{1+2b^{2}}$}
    \put(58.5,26){$\scriptstyle{1+b^{2}}$}
    \put(69,26){$\scriptstyle{-b^{2}}$}
    \put(45.5,16){$\scriptstyle{\boldsymbol{\alpha}_{1}}$}
    \put(45.5,33.5){$\scriptstyle{\boldsymbol{\alpha}_{2}}$}
     \put(54.6,27){$\scriptstyle{\boldsymbol{\alpha}_{3}}$}
     \put(64.6,27){$\scriptstyle{\boldsymbol{\alpha}_{4}}$}
     \put(74.6,27){$\scriptstyle{\boldsymbol{\alpha}_{5}}$}
      \put(88.6,27){$\scriptstyle{\boldsymbol{\alpha}_{N-2}}$}
     \put(98.6,27){$\scriptstyle{\boldsymbol{\alpha}_{N-1}}$}
  \end{picture}
\end{equation}
The vector $\boldsymbol{\rho}$ in \eqref{action} has been chosen in such a  way that the theory enjoys the conformal invariance. Explicitly it has the form
\begin{equation}
   \boldsymbol{\rho}=\frac{1}{2}\sum_{r=1}^{N-1}\hat{\boldsymbol{\alpha}}_{r},
\end{equation}
where $\hat{\boldsymbol{\alpha}}_{r}$ is the dual basis: $(\boldsymbol{\alpha}_{r},\hat{\boldsymbol{\alpha}}_{s})=\delta_{rs}$. This conformal field theory has the central charge
\begin{equation}\label{c-ON}
   c(N)=1+12(\boldsymbol{\rho},\boldsymbol{\rho})=\frac{(N-1)(x-N+3)(2x-N+2)}{2x(x+1)},\quad
   x=\begin{cases}
      b^{2}\quad\text{for}\quad N\in2\mathbb{Z}\\
      -1-b^{2}\quad\text{for}\quad N\in2\mathbb{Z}+1.
   \end{cases}
\end{equation}
The corresponding conformal algebra is generated by the stress-energy tensor and the spin $4$ field. Its explicit form is too complicated to be presented here.
This algebra coincides with the chiral local algebra of the coset CFT \cite{Goddard:1986ee}
\begin{equation}
   \frac{\hat{\mathfrak{so}}(N)_{k}}{\hat{\mathfrak{so}}(N-1)_{k}}\quad\text{with}\quad k=x-N+3.
\end{equation}

Consider the problem of computation of correlation functions in the theory \eqref{action}. For simplicity, we assume the geometry of the two-sphere. Using the well-known trick \cite{Goulian:1990qr} one can show that the $n-$point correlation function of the exponential fields
\begin{equation}
  \langle V_{\mathbf{a}_{1}}(\xi_{1},\bar{\xi}_{1})\dots V_{\mathbf{a}_{n}}(\xi_{n},\bar{\xi}_{n})\rangle,\quad
  \text{where}\quad
  V_{\mathbf{a}}=e^{(\mathbf{a},\varphi)},
\end{equation}
being considered as a function of the total charge $\mathbf{a}=\mathbf{a}_{1}+\dots+\mathbf{a}_{n}$ has multiple poles at the values
\begin{equation}\label{onshell-cond}
\mathbf{a}+\sum_{j=1}^{N-1}m_{j}\boldsymbol{\alpha}_{j}=2\boldsymbol{\rho},
\end{equation}
where $m_{j}$'s are some non-negative integer numbers. The multiple residues at these poles are proportional to the free-field correlation functions
\begin{equation}\label{GL-relation}
  \textrm{Res}\,\langle V_{\mathbf{a}_{1}}(\xi_{1},\bar{\xi}_{1})\dots V_{\mathbf{a}_{n}}(\xi_{n},\bar{\xi}_{n})\rangle
  \biggl|_{\mathbf{a}+\sum_{j=1}^{N-1}m_{j}\boldsymbol{\alpha}_{j}=2\boldsymbol{\rho}}=(-\Lambda)^{\sum_{j}m_{j}}
  \langle V_{\mathbf{a}_{1}}(\xi_{1},\bar{\xi}_{1})\dots V_{\mathbf{a}_{n}}(\xi_{n},\bar{\xi}_{n})
  \prod_{j=1}^{n}\frac{\left(\mathcal{S}_{j}\right)^{m_{j}}}{\pi^{m_{j}}m_{j}!}\rangle_{\textrm{\tiny{FF}}},
\end{equation}
where  $\mathcal{S}_{j}=\int e^{(\boldsymbol{\alpha}_{j},\varphi(\xi,\bar{\xi}))}d^{2}\xi$.  The multiple integral in the r.h.s. in \eqref{GL-relation} converges in the domain
\begin{equation}
   -1<b^{2}<0,\quad (\mathbf{a}_{k},\boldsymbol{\alpha}_{j})<1.
\end{equation}
Outside this domain this integral should be understood as an analytical continuation. A useful tool for analytical continuation is the well known identity for Coulomb integrals  \cite{Baseilhac:1998eq}
\begin{multline}\label{Fateev-integral}
 \int\mathcal{D}_{n}(x)\prod_{i=1}^{n}\prod_{j=1}^{n+m+2}|x_{i}-t_{j}|^{2p_{j}}\,d^{2}\vec{x}_{n}=
 \prod_{j=1}^{n+m+2}\gamma(1+p_{j})\prod_{i<j}|t_{i}-t_{j}|^{2+2p_{i}+2p_{j}}
 \times\\\times
 \int\mathcal{D}_{m}(y)\prod_{i=1}^{m}\prod_{j=1}^{n+m+2}|y_{i}-t_{j}|^{-2-2p_{j}}\,d^{2}\vec{y}_{m},
\end{multline}
where
\begin{equation*}
  \mathcal{D}_{n}(x)=\prod_{i<j}|x_{i}-x_{j}|^{2},\qquad
  d^{2}\vec{x}_{n}=\frac{1}{\pi^{n}n!}\prod_{j=1}^{n}d^{2}x_{j},\quad
  \gamma(x)=\frac{\Gamma(x)}{\Gamma(1-x)}\quad\text{and}\quad
  \sum_{j=1}^{n+m+2} p_{j}=-n-1.
\end{equation*}
This relation holds for any $n$ and $m$ and for all $-1<p_{j}<0$. The last condition is equivalent to the absence of singularity at infinity, which is always the case for correlation functions \eqref{GL-relation}.
It can be used for computation of many interesting Coulomb integrals appearing in CFT (see for example  \cite{Fateev:2007qn}). With the help of the relation \eqref{Fateev-integral} we can massage the free-field correlation function in the r.h.s. of \eqref{GL-relation}. It is convenient to define operators $\mathcal{R}_{r}$ which act on a space of free-field correlation functions as follows
\begin{multline}
  \langle V_{\mathbf{a}_{1}}(\xi_{1},\bar{\xi}_{1})\dots V_{\mathbf{a}_{n}}(\xi_{n},\bar{\xi}_{n})
  \prod_{j=1}^{n}\frac{\left(\mathcal{S}_{j}\right)^{m_{j}}}{\pi^{m_{j}}m_{j}!}\rangle_{\textrm{\tiny{FF}}}\overset{\mathcal{R}_{r}}{\longrightarrow}\\
  \overset{\mathcal{R}_{r}}{\longrightarrow}
  \prod_{s:(\boldsymbol{\alpha}_{r},\boldsymbol{\alpha}_{s})\neq0}\left(\frac{1}{\gamma\bigl((\boldsymbol{\alpha}_{r},\boldsymbol{\alpha}_{s})\bigr)}\right)^{m_{s}}
  \prod_{i=1}^{n}\mathcal{N}_{r}(\mathbf{a}_{i})\langle V_{\mathbf{a}_{1}+\boldsymbol{\alpha}_{r}}(\xi_{1},\bar{\xi}_{1})\dots V_{\mathbf{a}_{n}+\boldsymbol{\alpha}_{r}}(\xi_{n},\bar{\xi}_{n})
  \prod_{j=1}^{n}\frac{\left(\mathcal{\tilde{S}}_{j}\right)^{\tilde{m}_{j}}}{\pi^{\tilde{m}_{j}}\tilde{m}_{j}!}\rangle_{\textrm{\tiny{FF}}},
\end{multline}
where
\begin{equation}
    \mathcal{N}_{r}(\mathbf{a})=\frac{1}{\gamma\bigl((\boldsymbol{\alpha}_{r},\mathbf{a})\bigr)},\qquad
    \tilde{\mathcal{S}}_{j}=\int e^{(\tilde{\boldsymbol{\alpha}_{j}},\varphi(\xi,\bar{\xi}))}d^{2}\xi\quad\text{with}\quad
    \tilde{\boldsymbol{\alpha}}_{j}=
   \begin{cases}
   -\boldsymbol{\alpha}_{j}\quad\text{if}\quad j=r,\\
   \boldsymbol{\alpha}_{j}+\boldsymbol{\alpha}_{r}\quad\text{if}\quad (\boldsymbol{\alpha}_{j},\boldsymbol{\alpha}_{r})\neq0\\
   \boldsymbol{\alpha}_{j}\quad\text{otherwize}
   \end{cases}
\end{equation}
and
\begin{equation}
   \tilde{m}_{r}=n+\sum_{s:(\boldsymbol{\alpha}_{r},\boldsymbol{\alpha}_{s})\neq0}m_{s}-m_{r}-2,\quad \tilde{m}_{s}=m_{s}\quad\text{for}\quad s\neq r.
\end{equation}
We note that the operator $\mathcal{R}_{r}$ is just an application of the integral identity \eqref{Fateev-integral} to the contribution of the fermionic root $\boldsymbol{\alpha}_{r}$. It changes the roots $\boldsymbol{\alpha}_{s}\rightarrow {\tilde{\boldsymbol{\alpha}}_{s}}$ as well as shifts the charges of the exponential fields according to
\begin{equation}
    \mathbf{a}\rightarrow\mathbf{a}+\boldsymbol{\alpha}_{r}.
\end{equation}
As explained in \cite{Litvinov:2016mgi}, the operator $\mathcal{R}_{r}$ can be lifted to the operator, called fermionic reflection operator, acting on the total  ``off-shell'' correlation functions. Therefore  it serves as an isomorphism between different conformal field theories, corresponding to different root systems.  In particular, it establishes an isomorphism between $W-$algebras in different realizations.  Applying different $\mathcal{R}_{r}$'s one can reduce the number of fermionic roots. Actually, if $\boldsymbol{\alpha}_{s}$ ($s\neq r$) is a fermionic root then the transformed root $\tilde{\boldsymbol{\alpha}}_{s}$ is a bosonic one (i.e. $(\tilde{\boldsymbol{\alpha}}_{s},\tilde{\boldsymbol{\alpha}}_{s})\neq-1$) and vice versa. It will be convenient to draw bosonic roots as circles an use conventions similar to the Dynkin ones. Namely,
\begin{equation*}
\hspace*{-4cm}
\begin{picture}(200,40)(215,123)
    \Thicklines
    \unitlength 5pt
    \put(69,25){\circle{2}}
    \put(70,25){\line(1,0){5}}
    \put(63,25){\line(1,0){5}}
    \put(60,25){\circle{.2}}
    \put(61,25){\circle{.2}}
    \put(62,25){\circle{.2}}
    \put(77,25){\circle{.2}}
    \put(78,25){\circle{.2}}
    \put(76,25){\circle{.2}}
    \put(77,25){\circle{.2}}
    \put(78,25){\circle{.2}}
    \put(64.5,26){$-x$}
    \put(70.5,26){$-x$}
\end{picture}
\hspace*{-.5cm}
\sim\;\;
\begin{pmatrix}
    \dots&-x&0\\
    -x&2x&-x\\
    0&-x&\dots
    \end{pmatrix},\;
\begin{picture}(200,40)(215,123)
   \Thicklines
    \unitlength 5pt
   \put(45,25){\circle{2}}
    \put(45.9,24.6){\line(1,0){8.1}}
    \put(45.9,25.4){\line(1,0){8.1}}
    \put(50.5,25){\line(-1,1){1}}
    \put(50.5,25){\line(-1,-1){1}}
    \put(55,25){\circle{2}}
    \put(47.5,27){$-2x$}
  \end{picture}•
\hspace*{-4.2cm}
\sim\;\;
\begin{pmatrix}
  4x&-2x\\
  -2x&2x
\end{pmatrix}
\;\;\text{etc}
\end{equation*}
One can show, that starting from the root system corresponding to the diagram \eqref{Dn} and applying successively transformation $\mathcal{R}_{r}$  one can reduce it to the diagram with one fermionic root and $N-2$ bosonic ones. For generic value of $N$ there are only two possibilities:
\begin{itemize}
\item
First, corresponds to the operator
\begin{equation}
   \mathcal{R}_{I}\overset{\text{def}}{=}\dots\left(\mathcal{R}_{6}\mathcal{R}_{7}\mathcal{R}_{8}\mathcal{R}_{9}\right)\left(\mathcal{R}_{5}\mathcal{R}_{6}\mathcal{R}_{7}\right)
   \left(\mathcal{R}_{4}\mathcal{R}_{5}\right)\left(\mathcal{R}_{3}\right)
\end{equation}
This transformation sends \eqref{Dn} to the coupled root system of $D$ and $A$ type
\begin{equation}\label{Dn-transformed}
\begin{picture}(300,100)(280,80)
    \Thicklines
    \unitlength 5pt
    \put(48,32){\circle{2}}
    \put(48,18){\circle{2}}
    \put(54.3,24,3){\line(-1,-1){5.5}}
    \put(54.3,25,7){\line(-1,1){5.5}}
    \put(55,25){\circle{2}}
    \put(66,25){\line(1,0){8}}
    \put(56,25){\line(1,0){8}}
    \put(65,25){\circle{2}}
    \put(75,25){\circle{2}}
    \put(76,25){\line(1,0){2}}
    \put(80,25){\circle{.2}}
    \put(81,25){\circle{.2}}
    \put(82,25){\circle{.2}}
    \put(84,25){\line(1,0){5}}
    \put(90,25){\circle{2}}
    \put(89.4,24,4){\line(1,1){1.2}}
    \put(89.4,25,6){\line(1,-1){1.2}}
    \put(91,25){\line(1,0){8}}
    \put(100,25){\circle{2}}
    \put(101,25){\line(1,0){8}}
    \put(110,25){\circle{2}}
    \put(111,25){\line(1,0){2}}
    \put(115,25){\circle{.2}}
    \put(116,25){\circle{.2}}
    \put(117,25){\circle{.2}}
    \put(45.5,15.5){$\scriptstyle{\boldsymbol{\alpha}_{1}}+\scriptstyle{\boldsymbol{\alpha}_{3}}$}
    \put(45.5,33.5){$\scriptstyle{\boldsymbol{\alpha}_{2}}+\scriptstyle{\boldsymbol{\alpha}_{3}}$}
     \put(53.5,27){$\scriptstyle{\boldsymbol{\alpha}_{4}}+\scriptstyle{\boldsymbol{\alpha}_{5}}$}
     \put(62.3,27){$\scriptstyle{\boldsymbol{\alpha}_{6}}+\scriptstyle{\boldsymbol{\alpha}_{7}}$}
     \put(72.3,27){$\scriptstyle{\boldsymbol{\alpha}_{8}}+\scriptstyle{\boldsymbol{\alpha}_{9}}$}
      \put(89.3,27){$\scriptstyle{\boldsymbol{\alpha}_{0}}$}
       \put(93.3,26){$\scriptstyle{-b^{2}}$}
       \put(85,26){$\scriptstyle{1+b^{2}}$}
     \put(97.3,27){$\scriptstyle{\boldsymbol{\alpha}_{3}}+\scriptstyle{\boldsymbol{\alpha}_{4}}$}
     \put(107.3,27){$\scriptstyle{\boldsymbol{\alpha}_{5}}+\scriptstyle{\boldsymbol{\alpha}_{6}}$}
  \end{picture}
\end{equation}
where
\begin{equation}
   \boldsymbol{\alpha}_{0}=
   \begin{cases}
      -(\boldsymbol{\alpha}_{3}+\dots+\boldsymbol{\alpha}_{N-1}),\quad\text{for}\quad N=2n,\\
      -(\boldsymbol{\alpha}_{3}+\dots+\boldsymbol{\alpha}_{N-2}),\quad\text{for}\quad N=2n+1.
   \end{cases}
\end{equation}
\item
Another corresponds to the operator
\begin{equation}\label{RII}
   \mathcal{R}_{II}\overset{\text{def}}{=}\dots\left(\mathcal{R}_{5}\mathcal{R}_{6}\mathcal{R}_{7}\mathcal{R}_{8}\right)\left(\mathcal{R}_{4}\mathcal{R}_{5}\mathcal{R}_{6}\right)
   \left(\mathcal{R}_{3}\mathcal{R}_{4}\right)\left(\mathcal{R}_{2}\right)
\end{equation}
and sends the system to the $C$ and $A$ coupled system
\begin{equation}\label{Dn-transformed-2}
\begin{picture}(300,100)(280,80)
    \Thicklines
    \unitlength 5pt
    \put(45,25){\circle{2}}
    \put(45.9,24.6){\line(1,0){8.1}}
    \put(45.9,25.4){\line(1,0){8.1}}
    \put(50.5,25){\line(-1,1){1}}
    \put(50.5,25){\line(-1,-1){1}}
    \put(55,25){\circle{2}}
    \put(66,25){\line(1,0){8}}
    \put(56,25){\line(1,0){8}}
    \put(65,25){\circle{2}}
    \put(75,25){\circle{2}}
    \put(76,25){\line(1,0){2}}
    \put(80,25){\circle{.2}}
    \put(81,25){\circle{.2}}
    \put(82,25){\circle{.2}}
    \put(84,25){\line(1,0){5}}
    \put(90,25){\circle{2}}
    \put(89.4,24,4){\line(1,1){1.2}}
    \put(89.4,25,6){\line(1,-1){1.2}}
    \put(91,25){\line(1,0){8}}
    \put(100,25){\circle{2}}
    \put(101,25){\line(1,0){8}}
    \put(110,25){\circle{2}}
     \put(111,25){\line(1,0){2}}
    \put(115,25){\circle{.2}}
    \put(116,25){\circle{.2}}
    \put(117,25){\circle{.2}}
     \put(42.3,27){$\scriptstyle{\boldsymbol{\alpha}_{1}}+\scriptstyle{\boldsymbol{\alpha}_{2}}$}
     \put(52.3,27){$\scriptstyle{\boldsymbol{\alpha}_{3}}+\scriptstyle{\boldsymbol{\alpha}_{4}}$}
     \put(62.3,27){$\scriptstyle{\boldsymbol{\alpha}_{5}}+\scriptstyle{\boldsymbol{\alpha}_{6}}$}
     \put(72.3,27){$\scriptstyle{\boldsymbol{\alpha}_{7}}+\scriptstyle{\boldsymbol{\alpha}_{8}}$}
      \put(89.3,27){$\scriptstyle{\boldsymbol{\alpha}_{0}}$}
      \put(93.3,26){$\scriptstyle{1+b^{2}}$}
       \put(85,26){$\scriptstyle{-b^{2}}$}
     \put(97.3,27){$\scriptstyle{\boldsymbol{\alpha}_{2}}+\scriptstyle{\boldsymbol{\alpha}_{3}}$}
     \put(107.3,27){$\scriptstyle{\boldsymbol{\alpha}_{4}}+\scriptstyle{\boldsymbol{\alpha}_{5}}$}
  \end{picture}
\end{equation}
where
\begin{equation}
   \boldsymbol{\alpha}_{0}=
   \begin{cases}
      -(\boldsymbol{\alpha}_{2}+\dots+\boldsymbol{\alpha}_{N-1}),\quad\text{for}\quad N=2n+1,\\
      -(\boldsymbol{\alpha}_{3}+\dots+\boldsymbol{\alpha}_{N-2}),\quad\text{for}\quad N=2n.
   \end{cases}
\end{equation}
\end{itemize}
The operator $\mathcal{R}_{r}$ establishes the relation between different correlation functions and provides an isomorphism between $W$-algebras in different realizations. From the considerations above, it is natural to conjecture that the corresponding $W-$algebra coincides with
\begin{equation}
   W(D(n+1|n))\quad\text{for}\quad N=2n+2,\quad
   W(D(n|n))\quad\text{for}\quad N=2n+1,
\end{equation}
where $D(m|n)\sim OSP(2m|2n)$ is the corresponding superalgebra. We note that the case $N=4$ corresponds to the superalgebra $D(2|1)$ which is exceptional since it admits one parametric deformation $D(2|1)\rightarrow D(2|1,\alpha)$. On the sigma model side it corresponds to the sigma model with $S^{3}=SU(2)$ target space which is a group manifold and hence it admits  a two-parametric integrable deformation due to Klimcik \cite{Klimcik:2014bta}.  See ref \cite{Bazhanov:2013cua} for more details on this model, its dual description and relation to the superalgebra $D(2|1,\alpha)$.
\section{Ricci flow equations}\label{ricci}
In section \ref{dual-L} we introduced the  theory \eqref{Lagrangian}, \eqref{Lagrangian2}, \eqref{Lagrangian3} and claimed that it possesses infinitely many local integrals of motion. In the leading order in the parameter $\Lambda$ we constructed first non-trivial conserved quantity (see appendix \ref{IM}). In this section, we formulate the dual sigma model description of the same theory.

The duality description emerges from the following fact \cite{Litvinov:2016mgi}. Consider two exponential operators $e^{(\boldsymbol{\alpha}_{r},\varphi)}$, $r=1,2$, such that
$(\boldsymbol{\alpha}_{r},\boldsymbol{\alpha}_{r})=-1$ and  $(\boldsymbol{\alpha}_{1},\boldsymbol{\alpha}_{2})\neq0$. And suppose they commute with integrable system $\{\mathbf{I}_{2k-1}\}$: $\mathbf{I}_{2k-1}=\int G_{2k}(z)dz$.  It means that
\begin{equation}
   \oint_{\mathcal{C}_{z}}G_{2k}(z)e^{(\boldsymbol{\alpha}_{r},\varphi(w))}dw=\partial_{z}\mathcal{V}_{k}^{(r)}(z),
\end{equation}
with some local field $\mathcal{V}_{k}^{(r)}(z)$. Then, one can explicitly check that there also exists a chiral field
\begin{equation}\label{sigma screening}
    \mathcal{V}_{1,2}=(\boldsymbol{\alpha}_{1},\partial\varphi)e^{(\boldsymbol{\beta}_{12},\varphi)},\quad\text{where}\quad
    \boldsymbol{\beta}_{12}=\frac{2}{(\boldsymbol{\alpha}_{1}+\boldsymbol{\alpha}_{2})^{2}}(\boldsymbol{\alpha}_{1}+\boldsymbol{\alpha}_{2}),
\end{equation}
such that
\begin{equation}
   \oint_{\mathcal{C}_{z}}W_{2k}(z)\mathcal{V}_{1,2}(w)dw=\partial_{z}\mathcal{V}_{k}^{(1,2)}(z),
\end{equation}
with some local $\mathcal{V}_{k}^{(1,2)}(z)$. This fact can be checked by explicit computation. We note that $\mathcal{V}_{1,2}(w)$  is defined modulo total derivative, i.e. up to a shift $\boldsymbol{\alpha}_{1}\rightarrow\boldsymbol{\alpha}_{1}+\zeta\boldsymbol{\beta}_{12}$. For our theory the fields $\mathcal{V}_{i,j}$ can be combined into two groups
\begin{equation}
\begin{aligned}
  &\mathbf{I}:\qquad&&\left(\mathcal{V}_{1,2},\mathcal{V}_{3,4},\mathcal{V}_{5,6},\dots\right),\\
  &\mathbf{II}:\qquad&&\left(\mathcal{V}_{3,1},\mathcal{V}_{3,2},\mathcal{V}_{4,5},\mathcal{V}_{6,7},\dots\right).
\end{aligned}
\end{equation}
Using the coordinate representation \eqref{vectors} and \eqref{vectors-2} we find that the corresponding exponents have the form
\begin{equation}
\begin{aligned}
  &(\boldsymbol{\beta}_{12},\boldsymbol{\beta}_{34},\boldsymbol{\beta}_{56},\dots)=\frac{1}{b}
  (E_{1},E_{2}-E_{1},E_{3}-E_{2},\dots),\\
  &(\boldsymbol{\beta}_{31},\boldsymbol{\beta}_{32},\boldsymbol{\beta}_{45},\dots)=\frac{i}{\sqrt{1+b^{2}}}
  (-e_{1}-e_{2},e_{1}-e_{2},e_{2}-e_{3},\dots).
\end{aligned}
\end{equation}
We note that groups $I$ and $II$ have mutually commuting exponents. Moreover, in the sigma model regime $b\in \mathbb R$ and $b\gg1$ the exponents in the group $I$ are real and the exponents of the group $II$ are purely imaginary. The Gram matrices for the group $I$ are proportional to
\begin{equation}
    \begin{aligned}
       &\text{Gram matrix for affine}\;B(n)\quad\text{for}\quad N=2n+1,\\
       &\text{Gram matrix for affine}\;C^{\vee}(n)\quad\text{for}\quad N=2n+2.
    \end{aligned}
\end{equation}

At this point, it is useful to recall the sausage model. As it was discussed in the introduction, in the deep UV the sausage looks like a very long cylinder with ends on both sides. Therefore, it is natural to expect the metric to be only slightly perturbed from being flat if we are not too close to the edge of the sausage. A careful inspection of the operator \eqref{sigma screening} (together with its anti-holomorphic counterpart) suggests that exactly this operators are natural candidates for this perturbation from the flat metric\footnote{It is also useful to make an analogy with string theory, where the operators \eqref{sigma screening} correspond to gravitons, i.e. fields that perturb target-space metrics.}. The same reasoning makes it obvious that these operators only correspond to the first linear perturbation and one has to include additional corrections in order to describe the region near the edge of the sausage. Also, one could worry that higher-order curvature correction in the renormalization of the metrics will become important near the cups of the sausage. However, as we will see, these corrections are suppressed in the limit $b\rightarrow\infty$.

Keeping in mind the considerations above, we consider the theory
\begin{equation}\label{SM-Lagrangian}
  \mathcal{L}=\frac{1}{8\pi}\bigl(\partial_{\mu}\varphi\bigr)^{2}+
  \Lambda\sum_{(i,j)\in\mathbf{I}}(\boldsymbol{\alpha}_{i},\partial\varphi)(\boldsymbol{\alpha}_{i},\bar{\partial}\varphi)e^{(\boldsymbol{\beta}_{ij},\varphi)}+\dots,
\end{equation}
which might be the dual sigma model description of the original theory \eqref{Lagrangian},\eqref{Lagrangian2}, \eqref{Lagrangian3}. By $\dots$ we mean possible counterterms. We note that the theory \eqref{SM-Lagrangian} has $\mathbf{P}$ symmetry, but lacks $\mathbf{C}$ symmetry for $N>4$. As we will see below, by performing the $T-$duality one can make the metric real, but introduce pure imaginary $B-$field, exactly as follows from the general action \eqref{Coset-action-deformed}. Now we come to the important point. We note that the theory \eqref{SM-Lagrangian} is non-renormalizable in a strong sense. One has to add infinite number of counter terms to ensure renormalizability. Let us study this question in the one-loop approximation. The parameter $b$ plays the role of the coupling constant in our theory. In order to probe the semiclassical region, we rescale
\begin{equation}
   \varphi=b X
\end{equation}
and consider the limit $b\rightarrow\infty$. Introducing standard notation $\alpha'=\frac{2}{b^{2}}$, we rewrite  \eqref{SM-Lagrangian} as a series in $\alpha'$
\begin{equation}
  \mathcal{L}=\frac{1}{4\pi\alpha'}G_{\mu\nu}(X)\partial_{a}X^{\mu}\partial_{a}X^{\nu}+O(1)\quad\text{at}\quad \alpha'\rightarrow0,
\end{equation}
where
\begin{equation}\label{metric-UV-asympt}
   G_{\mu\nu}(X)=\delta_{\mu\nu}+\Lambda\sum_{(i,j)\in\mathbf{I}}a_{i,\mu}a_{i,\nu}e^{(b_{ij},X)}+O(\Lambda^{2}),
\end{equation}
with the vectors $a_{i}$ and $b_{ij}$ being defined through the asymptotic
\begin{equation}
   a_{r}\overset{\text{def}}{=}\lim_{b\rightarrow\infty}b^{-1}\boldsymbol{\alpha}_{r},\qquad
   b_{ij}\overset{\text{def}}{=}\lim_{b\rightarrow\infty}b\boldsymbol{\beta}_{ij}+\dots.
\end{equation}
The subleading terms in the expansion \eqref{metric-UV-asympt} has to be chosen to ensure  the one-loop renormalizability of the theory. This condition leads to the RG group flow equation\footnote{For simplicity we've made the scale transformation $t\rightarrow t/\alpha'$.}
\begin{equation}\label{RG-equation-3}
  R_{\mu\nu}+2\nabla_{\mu}\nabla_{\nu}\Psi=-\dot{G}_{\mu\nu},
\end{equation}
where dot corresponds to the derivative with respect to the RG time $t$ which is proportional to the $\log(\Lambda_{UV})$. The function $\Psi$ in this equation  is more or less arbitrary and  can be chosen at will.  It describes the effect of possible RG-time dependent gradient diffeomorphisms\footnote{In general, diffeomorphism term in Ricci flow equation has the form $\nabla_{\mu}V_{\nu}+\nabla_{\nu}V_{\mu}$.}. It means that the solution to \eqref{RG-equation-3} is always a pair: the flowing metric  and the coordinate frame. In this sense, $\Psi$ is a gauge. There are at least two gauges used in the literature. First, is the so-called Hamilton gauge, in which $\Psi=0$. Another is a Friedan's gauge, such that $\Psi$ satisfied additional constraint 
\begin{equation}
    c_{0}+|\nabla \Psi|^{2}-\frac{1}{2}\Delta \Psi=-\dot{\Psi},
\end{equation}
where $c_{0}$ is a constant, which can always been set to zero by a linear shift of the function $\Psi\rightarrow\Psi-c_{0}t$. In this frame $\Psi$ can be treated as a dilaton field.  We find it convenient not to use any of these gauges, but rather properly adjust $\Psi$ in order to reduce the non-linearity of the equation \eqref{RG-equation-3}.

Now we proceed to the problem of our interest. Namely, we are looking for the solution to \eqref{RG-equation-3} with the UV asymptotic \eqref{metric-UV-asympt} prescribed by the bare action \eqref{SM-Lagrangian}. As we will see, one can use the gauge freedom to set the determinant of the metric to a constant. Below we will study in details two cases: $N=5$ and $N=6$. We will show that after a proper change of variables and $T-$dualities the corresponding solutions coincide with the ones following from the general action \eqref{Coset-action-deformed}.
\subsection{Metric for the deformed $O(5)$ model}
In this case, the target space is four-dimensional. The model corresponds to the graph
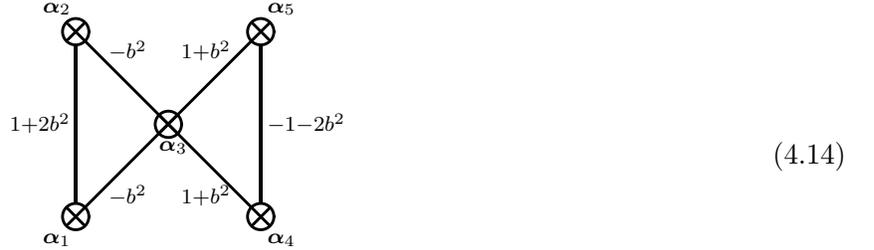
\begin{equation}\label{D2}
\begin{picture}(300,60)(140,110)
    \Thicklines
    \unitlength 5pt
    \put(48,32){\circle{2}}
    \put(48,18){\circle{2}}
    \put(54.4,24,4){\line(-1,-1){7}}
    \put(54.4,25,6){\line(-1,1){7}}
    \put(47.4,31.4){\line(1,1){1.2}}
    \put(47.4,18.6){\line(1,-1){1.2}}
    \put(48,19){\line(0,1){12}}
    \put(55,25){\circle{2}}
    \put(54.4,24,4){\line(1,1){1.2}}
    \put(54.4,25,6){\line(1,-1){1.2}}
    \put(54.4,24,4){\line(1,1){8.2}}
    \put(54.4,25,6){\line(1,-1){8.2}}
    \put(62,32){\circle{2}}
    \put(62,18){\circle{2}}
    \put(61.4,32.6){\line(1,-1){1.2}}
    \put(61.4,17.4){\line(1,1){1.2}}
    \put(62,19){\line(0,1){12}}
    \put(50.5,19){$\scriptstyle{-b^{2}}$}
    \put(50.5,30){$\scriptstyle{-b^{2}}$}
    \put(45.5,16){$\scriptstyle{\boldsymbol{\alpha}_{1}}$}
    \put(45.5,33.5){$\scriptstyle{\boldsymbol{\alpha}_{2}}$}
    \put(54.3,23){$\scriptstyle{\boldsymbol{\alpha}_{3}}$}
    \put(62.5,16){$\scriptstyle{\boldsymbol{\alpha}_{4}}$}
    \put(62.5,33.5){$\scriptstyle{\boldsymbol{\alpha}_{5}}$}
    \put(56,19){$\scriptstyle{1+b^{2}}$}
    \put(56,30){$\scriptstyle{1+b^{2}}$}
    \put(43,24.5){$\scriptstyle{1+2b^{2}}$}
    \put(62.5,24.5){$\scriptstyle{-1-2b^{2}}$}
  \end{picture}
  \vspace*{1cm}
\end{equation}
We choose the parametrization of the  vectors $\boldsymbol{\alpha}_{r}$ as in \eqref{vectors} (where $\beta=\sqrt{1+b^{2}}$)
\begin{equation}
    \boldsymbol{\alpha}_{1}=bE_{1}+i\beta e_{1},\;\boldsymbol{\alpha}_{2}=bE_{1}-i\beta e_{1},\;
    \boldsymbol{\alpha}_{3}=-bE_{1}+i\beta e_{2},\;
    \boldsymbol{\alpha}_{4}=bE_{2}-i\beta e_{2},\;\boldsymbol{\alpha}_{5}=-bE_{2}-i\beta e_{2},
\end{equation}
where $E_{1},E_{2},e_{1},e_{2}$ is the orthonormal basis in  $\mathbb{R}^{4}$ with coordinates $(x_{1},x_{2},x_{3},x_{4})$. We expect that  solution to  \eqref{RG-equation-3} should behave in the UV $t\rightarrow-\infty$ as
\begin{equation}\label{anzatz}
  G_{\mu\nu}=\delta_{\mu\nu}+e^{\alpha t}\left(A_{\mu\nu}e^{x_{1}}+B_{\mu\nu}e^{-x_{1}-x_{2}}+C_{\mu\nu}e^{-x_{1}+x_{2}}\right)+\dots,\quad
  \Psi=(\rho,x)+\dots,
\end{equation}
where
\begin{equation*}
  A_{\mu\nu}=\begin{pmatrix}
  1&0&i&0\\
  0&0&0&0\\
  i&0&-1&0\\
  0&0&0&0
  \end{pmatrix},\quad
  B_{\mu\nu}=\begin{pmatrix}
  0&0&0&0\\
  0&1&0&i\\
  0&0&0&0\\
  0&i&0&-1
  \end{pmatrix},\quad
  C_{\mu\nu}=\begin{pmatrix}
  0&0&0&0\\
  0&1&0&-i\\
  0&0&0&0\\
  0&-i&0&-1
  \end{pmatrix},
\end{equation*}
and $\rho$ is an unknown constant vector.   As explained above, the anzatz \eqref{anzatz} corresponds to the perturbation of the free theory by the operators
\begin{equation}
  \int(\boldsymbol{\alpha}_{1},\partial\varphi)(\boldsymbol{\alpha}_{1},\bar{\partial}\varphi)e^{(\boldsymbol{\beta}_{12},\varphi)}d^{2}z,\quad
  \int(\boldsymbol{\alpha}_{4},\partial\varphi)(\boldsymbol{\alpha}_{4},\bar{\partial}\varphi)e^{(\boldsymbol{\beta}_{34},\varphi)}d^{2}z,\quad
  \int(\boldsymbol{\alpha}_{5},\partial\varphi)(\boldsymbol{\alpha}_{5},\bar{\partial}\varphi)e^{(\boldsymbol{\beta}_{35},\varphi)}d^{2}z.
\end{equation}
Moreover, we expect that the terms shown by $\dots$ in \eqref{anzatz} are homogeneous polynomials in
\begin{equation}
  X=e^{\alpha t}e^{x_{1}},\quad Y=e^{\alpha t}e^{-x_{1}-x_{2}},\quad Z=e^{\alpha t}e^{-x_{1}+x_{2}},
\end{equation}
of degree $\geq2$. At the leading order we have from \eqref{RG-equation-3}
\begin{equation}\label{leading-order-equations}
   \alpha=\frac{3}{4},\qquad\rho=\left(-\frac{1}{4},0,-\frac{3i}{4},\frac{i}{2}\right).
\end{equation}
Solving the asymptotic problem one can see that the solution stays within the following anzatz for the metric $G_{\mu\nu}$ and $\Psi$
\begin{equation}\label{metric-anzatz}
   G_{\mu\nu}=
   \begin{pmatrix}
  F_{1}&0&iF_{5}&0\\
  0&F_{2}-\cosh(x_{2})F_{6}&0&-i\sinh(x_{2})F_{6}\\
  iF_{5}&0&F_{3}&0\\
  0&-i\sinh(x_{2})F_{6}&0&F_{4}+\cosh(x_{2})F_{6}
  \end{pmatrix},\qquad
  \Psi=(\rho,x)+F_{7}.
\end{equation}
In \eqref{metric-anzatz} the functions $F_{k}$ depend on $x_{1}$ and $t$ only: $F_{k}=F_{k}(x_{1},t)$. Moreover, equations \eqref{RG-equation-3} are compatible provided that
\begin{equation}\label{constraint-1}
   F_{2}=F_{4},\qquad F_{2}^{2}=1+F_{6}^{2}.
\end{equation}
The function $F_{7}$ is arbitrary and corresponds to the choice of gauge. We can choose it to ensure the additional relation
\begin{equation}\label{constraint-2}
   \det G=F_{1}F_{3}+F_{5}^{2}=1.
\end{equation}
With this choice the problem \eqref{RG-equation-3} has a unique solution satisfying  the asymptotic condition \eqref{anzatz}. It is given by
\begin{equation}\label{solution}
  \begin{gathered}
   F_{3}=\frac{(1-U)(1-UV)}{1-U^{2}V},\qquad F_{5}=\frac{U(1-V)}{1-U^{2}V}+\frac{2}{3}\frac{UV(1-U)}{1-U^{2}V},\\
   F_{6}=V^{\frac{1}{2}}\left(\frac{(1+V)}{(1-V)^{2}}\frac{1+U^{2}V}{1-U^{2}V}-\left(\frac{1}{2}+\frac{4V}{(1-V)^{2}}\right)\frac{U}{1-U^{2}V}\right),\quad
   F_{7}=\log\left(\frac{(1-UV)^{2}}{1-V}\right)
  \end{gathered}
\end{equation}
where
\begin{equation*}
   U=e^{\frac{3t}{4}}e^{x_{1}},\qquad V=\frac{1}{4}e^{\frac{3t}{2}}e^{-2x_{1}}.
\end{equation*}
The other functions $F_{k}$ are obtained from the constraints \eqref{constraint-1} and \eqref{constraint-2}.

There is a choice of coordinates, such that the metric $G_{\mu\nu}$ has a particularly simple form. Namely, we perform a shift
\begin{equation}\label{shift}
   x_{3}\rightarrow x_{3}+\frac{i}{3}\log\left((1-U)^{3}(1-UV)\right).
\end{equation}
This shift diagonalizes the $(1,3)$ part of the metric. Moreover, it is convenient to introduce new coordinates  $\zeta$, $\theta$, $\phi_{1}$, $\phi_{2}$ and the function $\kappa(t)$ by the following equations
\begin{equation}
   F_{3}=\frac{\kappa(1-\zeta^{2})}{(1-\kappa^{2}\zeta^{2})},\quad \tanh\left(\frac{x_{2}}{2}\right)=\sin\theta,\quad
   \kappa=\frac{2-e^{\frac{3t}{2}}}{2+e^{\frac{3t}{2}}},\quad \phi_{1}=\frac{x_{3}}{2},\quad \phi_{2}=\frac{x_{4}}{2}-\frac{i}{2}\log\cos\theta.
\end{equation}
In these coordinates the metric has the form (after rescaling $ds^{2}\rightarrow4\nu ds^{2}$, $t\rightarrow4\nu t+\log2$)
\begin{equation}\label{good-metric}
   ds^{2}=\frac{\kappa}{\nu}\left(\frac{d\zeta^{2}}{(1-\zeta^{2})(1-\kappa^{2}\zeta^{2})}+\frac{(1-\zeta^{2})d\phi_{1}^{2}}{(1-\kappa^{2}\zeta^{2})}+\zeta^{2}d\theta^{2}+
   2i\zeta^{2}\tan\theta d\theta d\phi_{2}+
   \frac{(1-\kappa^{2}\zeta^{4}\sin^{2}\theta)d\phi_{2}^{2}}{\kappa^{2}\zeta^{2}\cos^{2}\theta}\right).
\end{equation}
The metric \eqref{good-metric} can be found in \cite{Hoare:2015gda}. It satisfies Ricci flow equation \eqref{RG-equation-3} with the function $\Psi$ given by
\begin{equation}\label{S4-dilaton}
   \Psi=\frac{1}{2}\log\left(\frac{(1-\kappa^{2}\zeta^{2})^{2}}{\kappa(1-\kappa^{2})\zeta^{2}\cos^{2}\theta}\right)-i\phi_{2},
\end{equation}
and $\kappa=-\tanh(3\nu t)$. Moreover \eqref{S4-dilaton} satisfies
\begin{equation}
   |\nabla \Psi|^{2}-\frac{1}{2}\Delta\Psi=-\dot{\Psi},
\end{equation}
and hence $\Psi$ can be regarded as a dilaton field.  Now we perform the T-duality in the $\phi_{2}$ isometry direction making the metric diagonal
\begin{equation}\label{Klimcik-metric}
   d\tilde{s}^{2}=\frac{\kappa}{\nu}\Biggl[\frac{d\zeta^{2}}{(1-\zeta^{2})(1-\kappa^{2}\zeta^{2})}+\frac{(1-\zeta^{2})d\phi_{1}^{2}}{(1-\kappa^{2}\zeta^{2})}+
   \frac{\zeta^{2}}{1-\kappa^{2}\zeta^{4}\sin^{2}\theta}\left(d\theta^{2}+\cos^{2}\theta\, d\phi_{2}^{2}\right)\Biggr],
\end{equation}
but generating a non-zero pure imaginary $B-$field
\begin{equation}\label{Klimcik-Bfield}
   B=\frac{i\kappa^{2}\sin\theta\cos\theta\zeta^{4}}{\nu(1-\kappa^{2}\zeta^{4}\sin^{2}\theta)}d\theta\wedge d\phi_{2}.
\end{equation}
In this form the metric and the $B-$field are exactly the same as follows from the general action \eqref{Coset-action-deformed} with the choice of coordinates in the coset as in appendix \ref{group-element}. We note that  the metric \eqref{Klimcik-metric} and the $B-$field \eqref{Klimcik-Bfield} satisfy
\begin{equation}\label{Ricci-modified}
      R_{\mu\nu}-\frac{1}{4}H_{\mu}^{\lambda\sigma}H_{\nu\lambda\sigma}+\nabla_{\mu}V_{\nu}+\nabla_{\nu}V_{\mu}=-\dot{G}_{\mu\nu},\quad
      H_{\lambda\mu\nu}V^{\lambda}-\frac{1}{2}\nabla_{\lambda}H^{\lambda}_{\mu\nu}+\nabla_{\mu}\omega_{\nu}-\nabla_{\nu}\omega_{\nu}=-\dot{B}_{\mu\nu}
\end{equation}
where $V_{\mu}$ and $\omega=\omega_{\mu}dx^{\mu}$ are given by
\begin{equation*}
  V_{\mu}=\left(\kappa ^2 \xi  \left(\frac{2 \xi ^2 \sin ^2(\theta )}{1-\kappa ^2 \xi ^4 \sin ^2(\theta )}-\frac{2}{1-\kappa ^2 \xi ^2}\right),0,0,0\right),\qquad
  \omega=\frac{i\kappa\zeta^{2}\cos^{2}\theta}{(1-\kappa^{2}\zeta^{4}\sin^{2}\theta)}d\phi_{2}.
\end{equation*}
In the limit $\nu\rightarrow0$ the metric \eqref{Klimcik-metric} approaches the metric of the round four-sphere, while the $B-$field, the vector $V_{\mu}$ and the one-form $\omega$ vanish.
\subsection{Metric for the deformed $O(6)$ model}
The model in this case is five-dimensional and defined by the  graph
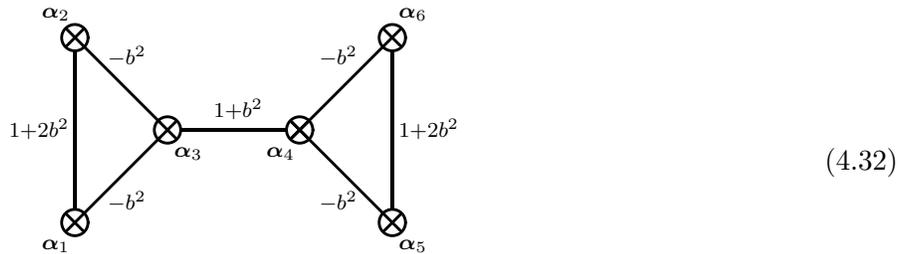
\begin{equation}\label{D3}
\begin{picture}(300,60)(160,110)
    \Thicklines
    \unitlength 5pt
    \put(48,32){\circle{2}}
    \put(48,18){\circle{2}}
    \put(54.4,24,4){\line(-1,-1){7}}
    \put(54.4,25,6){\line(-1,1){7}}
    \put(47.4,31.4){\line(1,1){1.2}}
    \put(47.4,18.6){\line(1,-1){1.2}}
    \put(48,19){\line(0,1){12}}
    \put(55,25){\circle{2}}
    \put(54.4,24,4){\line(1,1){1.2}}
    \put(54.4,25,6){\line(1,-1){1.2}}
    \put(56,25){\line(1,0){8}}
    \put(65,25){\circle{2}}
    \put(64.4,24,4){\line(1,1){1.2}}
    \put(64.4,25,6){\line(1,-1){1.2}}
    \put(64.4,24,4){\line(1,1){8.2}}
    \put(64.4,25,6){\line(1,-1){8.2}}
    \put(72,32){\circle{2}}
    \put(72,18){\circle{2}}
    \put(71.4,32.6){\line(1,-1){1.2}}
    \put(71.4,17.4){\line(1,1){1.2}}
    \put(72,19){\line(0,1){12}}
    \put(50.5,19){$\scriptstyle{-b^{2}}$}
    \put(50.5,30){$\scriptstyle{-b^{2}}$}
    \put(45.5,16){$\scriptstyle{\boldsymbol{\alpha}_{1}}$}
    \put(45.5,33.5){$\scriptstyle{\boldsymbol{\alpha}_{2}}$}
    \put(55.5,23){$\scriptstyle{\boldsymbol{\alpha}_{3}}$}
    \put(62.5,23){$\scriptstyle{\boldsymbol{\alpha}_{4}}$}
    \put(72.5,16){$\scriptstyle{\boldsymbol{\alpha}_{5}}$}
    \put(72.5,33.5){$\scriptstyle{\boldsymbol{\alpha}_{6}}$}
    \put(66.5,19){$\scriptstyle{-b^{2}}$}
    \put(66.5,30){$\scriptstyle{-b^{2}}$}
    \put(43,24.5){$\scriptstyle{1+2b^{2}}$}
    \put(72.5,24.5){$\scriptstyle{1+2b^{2}}$}
    \put(58.5,26){$\scriptstyle{1+b^{2}}$}
  \end{picture}
  \vspace*{1cm}
\end{equation}
The vectors $\boldsymbol{\alpha}_{r}$ can be parameterized as follows ($\beta=\sqrt{1+b^{2}}$)
\begin{equation}
\begin{aligned}
    &\boldsymbol{\alpha}_{1}=bE_{1}+i\beta e_{1},\quad&&\boldsymbol{\alpha}_{2}=bE_{1}-i\beta e_{1},\quad&&\boldsymbol{\alpha}_{3}=-bE_{1}+i\beta e_{2},\\
    &\boldsymbol{\alpha}_{4}=bE_{2}-i\beta e_{2},\quad&&\boldsymbol{\alpha}_{5}=-bE_{2}+i\beta e_{3},\quad&&\boldsymbol{\alpha}_{6}=-bE_{2}-i\beta e_{3},
\end{aligned}
\end{equation}
where $(E_{1},E_{2},e_{1},e_{2},e_{3})$ is an orthonormal basis in $\mathbb{R}^{5}$ with coordinates $(x_{1},x_{2},x_{3},x_{4},x_{5})$. In this case we are looking for the solution to the problem  \eqref{RG-equation-3} in the form
\begin{equation}\label{anzatz-2}
  G_{\mu\nu}=\delta_{\mu\nu}+2e^{\alpha t}\left(A_{\mu\nu}e^{x_{1}}+B_{\mu\nu}e^{-x_{2}}+C_{\mu\nu}e^{-x_{1}+x_{2}}\right)+\dots,\quad
  \Psi=(\rho,x)+\dots,
\end{equation}
where
\begin{equation*}
  A_{\mu\nu}=\begin{pmatrix}
  1&0&i&0&0\\
  0&0&0&0&0\\
  i&0&-1&0&0\\
  0&0&0&0&0\\
  0&0&0&0&0
  \end{pmatrix},\quad
  B_{\mu\nu}=\begin{pmatrix}
  0&0&0&0&0\\
  0&1&0&0&i\\
  0&0&0&0&0\\
  0&0&0&0&0\\
  0&i&0&0&-1
  \end{pmatrix},\quad
  C_{\mu\nu}=\begin{pmatrix}
  0&0&0&0&0\\
  0&1&0&-i&0\\
  0&0&0&0&0\\
  0&-i&0&-1&0\\
  0&0&0&0&0
  \end{pmatrix}.
\end{equation*}
It corresponds to the perturbation by the operators
\begin{equation}
  \int(\boldsymbol{\alpha}_{1},\partial\varphi)(\boldsymbol{\alpha}_{1},\bar{\partial}\varphi)e^{(\boldsymbol{\beta}_{12},\varphi)}d^{2}z,\quad
  \int(\boldsymbol{\alpha}_{4},\partial\varphi)(\boldsymbol{\alpha}_{4},\bar{\partial}\varphi)e^{(\boldsymbol{\beta}_{34},\varphi)}d^{2}z,\quad
  \int(\boldsymbol{\alpha}_{6},\partial\varphi)(\boldsymbol{\alpha}_{6},\bar{\partial}\varphi)e^{(\boldsymbol{\beta}_{56},\varphi)}d^{2}z.
\end{equation}
At the leading order we obtain
\begin{equation}
   \alpha=\frac{2}{3},\qquad \rho=\frac{1}{6}(-1,1,-4i,2i,4i).
\end{equation}
Then one can check that the solution stays within the following  anzatz
\begin{equation}\label{metric-anzatz-2}
  \begin{gathered}
   G_{\mu\nu}=
   \begin{pmatrix}
  F_{6}&\frac{1}{2}F_{5}&iF_{2}&-\frac{i}{2}F_{5}&\frac{i}{2}F_{5}\\
  \frac{1}{2}F_{5}&F_{1}+e^{-x_{2}}F_{3}+e^{x_{2}-x_{1}}F_{4}&0&-ie^{x_{2}-x_{1}}F_{4}&ie^{-x_{2}}F_{3}\\
  iF_{2}&0&F_{1}-F_{2}+F_{5}&0&0\\
  -\frac{i}{2}F_{5}&-ie^{x_{2}-x_{1}}F_{4}&0&F_{1}-e^{x_{2}-x_{1}}F_{4}&0\\
  \frac{i}{2}F_{5}&ie^{-x_{2}}F_{3}&0&0&F_{1}-e^{-x_{2}}F_{3}
  \end{pmatrix},\\
  \Psi=(\rho,x)+F_{7},
  \end{gathered}
\end{equation}
where the functions $F_{k}=F_{k}(x_{1},t)$ do not depend on $x_{2}$. The function $F_{7}$ in \eqref{metric-anzatz-2} is arbitrary. We use this freedom to set the determinant of the matrix $G$ to a $t-$dependent constant.  Now the solution is unique and explicitly given by (here $X=e^{\frac{2}{3}t}e^{x_{1}}$, $Y=e^{\frac{4}{3}t}e^{-x_{1}}$)
\begin{equation}
\begin{gathered}
     F_{1}=\frac{1-XY}{1+XY}+\frac{Y \left(1-XY\right)^{-\frac{1}{3}}}{2 \left(1+XY\right)},\;\;
   F_{2}=\frac{X\left(1-XY\right)^{\frac{1}{3}}}{\left(1+XY\right)},\;\;
   F_{3}=\frac{(XY)^{\frac{1}{3}}\sqrt{4 \left(1-XY\right)^{\frac{4}{3}}+2 Y}}{2 \left(1-XY\right)^{\frac{1}{3}}\left(1+XY\right)},\\
   F_{4}=\frac{(XY)^{\frac{1}{3}}\left(4\left(1-XY\right)-2X \left(1-XY\right)^{\frac{1}{3}}+\frac{Y}{\left(1-XY\right)^{\frac{1}{3}}}\right)}
   {2 \left(1+XY\right) \sqrt{4 \left(1-XY\right)^{\frac{4}{3}}+2 Y}},\;\;
   F_{5}=-\frac{Y \left(1-XY\right)^{-\frac{1}{3}}}{2 \left(1+XY\right)},\\
   F_{7}=\frac{3}{4} \log \left(\left(1-XY\right)^{\frac{4}{3}}+\frac{Y}{2}\right).
\end{gathered}
\end{equation}
and lengthy expression for $F_{6}$ which can be obtained from the constraint $ \det G=\frac{1-XY}{1+XY}$.
Now we change the coordinates ($\kappa=-\tanh t$)
\begin{equation}
\begin{gathered}
   x_{1}=\log\left(\frac{\zeta^{2}(-2\sinh t)^{\frac{2}{3}}}{2\cosh^{2}t(1-\kappa^{2}\zeta^{2})}\right),\quad
   x_{2}=\log\left(\frac{-(-2\sinh t)^{\frac{1}{3}}\zeta\tanh^{2}\theta}{\cosh t}\right)
   \\
   x_{3}=2\phi_{1}-i\log\left(\frac{1-\kappa^{2}\zeta^{2}}{1-\zeta^{2}}\right),\quad
   x_{4}=2\phi_{2}-i\log\left(\frac{\zeta}{\cos^{2}\theta}\right),\quad
   x_{5}=2\phi_{3}+i\log\left(\zeta\sin^{2}\theta\right)
\end{gathered}
\end{equation}
In these coordinates the metric has the form ($ds^{2}\rightarrow 4ds^{2}$, $t\rightarrow 4\nu t$)
\begin{multline}\label{good-metric-S5-2}
   ds^{2}=\frac{\kappa}{\nu}\left(\frac{d\zeta^{2}}{(1-\zeta^{2})(1-\kappa^{2}\zeta^{2})}+\frac{(1-\zeta^{2})d\phi_{1}^{2}}{(1-\kappa^{2}\zeta^{2})}+\zeta^{2}d\theta^{2}+
   \right.\\+\left.
   2i\zeta^{2}\tan\theta d\theta d\phi_{2}+
   \frac{(1-\kappa^{2}\zeta^{4}\sin^{2}\theta)d\phi_{2}^{2}}{\kappa^{2}\zeta^{2}\cos^{2}\theta}+
   \frac{1}{\kappa^{2}\zeta^{2}\sin^{2}\theta} d\phi_{3}^{2}\right).
\end{multline}
This metric is $T$-dual to the metric from \cite{Hoare:2015gda}
\begin{multline}\label{good-metric-S5}
   ds^{2}=\frac{\kappa}{\nu}\left(\frac{d\zeta^{2}}{(1-\zeta^{2})(1-\kappa^{2}\zeta^{2})}+\frac{(1-\zeta^{2})d\phi_{1}^{2}}{(1-\kappa^{2}\zeta^{2})}+\zeta^{2}d\theta^{2}+
   \right.\\+\left.
   2i\zeta^{2}\tan\theta d\theta d\phi_{2}+
   \frac{(1-\kappa^{2}\zeta^{4}\sin^{2}\theta)d\phi_{2}^{2}}{\kappa^{2}\zeta^{2}\cos^{2}\theta}+
   \zeta^{2}\sin^{2}\theta d\phi_{3}^{2}\right),
\end{multline}
in the $\phi_{3}$ isometry direction. The metric \eqref{good-metric-S5} satisfies  Ricci flow equations \eqref{RG-equation-3} with $\Psi$ and $\kappa$ given by
\begin{equation}
   \Psi=\frac{1}{2}\log\left(\frac{(1-\kappa^{2}\zeta^{2})^{3}}{\kappa(1-\kappa^{2})^{\frac{3}{2}}\zeta^{2}\cos^{2}\theta}\right)-2i\phi_{2},\qquad \kappa=-\tanh(4\nu t).
\end{equation}
We can now perform the $T$-duality in the $\phi_{2}$ isometry making the metric diagonal
\begin{equation}\label{Klimcik-metric-S5}
   d\tilde{s}^{2}=\frac{\kappa}{\nu}\Biggl[\frac{d\zeta^{2}}{(1-\zeta^{2})(1-\kappa^{2}\zeta^{2})}+\frac{(1-\zeta^{2})d\phi_{1}^{2}}{(1-\kappa^{2}\zeta^{2})}+
   \frac{\zeta^{2}}{1-\kappa^{2}\zeta^{4}\sin^{2}\theta}\left(d\theta^{2}+\cos^{2}\theta\, d\phi_{2}^{2}\right)+\zeta^{2}\sin^{2}\theta\, d\phi_{3}^{2}\Biggr],
\end{equation}
and generating the non-zero pure imaginary $B-$field
\begin{equation}\label{Klimcik-Bfield-S5}
   B=\frac{i\kappa^{2}\sin\theta\cos\theta\zeta^{4}}{\nu(1-\kappa^{2}\zeta^{4}\sin^{2}\theta)}d\theta\wedge d\phi_{2}.
\end{equation}
In this form the metric and the $B-$field coincide with the ones obtained from the deformed action \eqref{Coset-action-deformed} in \cite{Arutyunov:2013ega}. The metric \eqref{Klimcik-metric-S5} and the $B-$field \eqref{Klimcik-Bfield-S5} satisfy \eqref{Ricci-modified} with
\begin{equation*}
  V_{\mu}=\left(\kappa ^2 \xi  \left(\frac{2 \xi ^2 \sin ^2(\theta )}{1-\kappa ^2 \xi ^4 \sin ^2(\theta )}-\frac{3}{1-\kappa ^2 \xi ^2}\right),0,-\frac{\kappa ^2 \xi ^4 \sin (2 \theta )}{2 \left(1-\kappa ^2 \xi ^4 \sin ^2(\theta )\right)},0,0\right),\;
  \omega=\frac{i\kappa\zeta^{2}\cos^{2}\theta}{(1-\kappa^{2}\zeta^{4}\sin^{2}\theta)}d\phi_{2}.
\end{equation*}

At the end of this section we note, that the theory with $N=2n+1$ is self-dual with respect to the transformation $b^{2}\rightarrow-1-b^{2}$, while the theory with $N=2n+2$  is not. It is interesting to study the theory with $N=2n+2$ in this dual domain of parameters $b^{2}\rightarrow-1-b^{2}$. In particular, it will be interesting to construct the corresponding solution of Ricci flow equation and relate it to some deformed coset sigma model. For interested reader, we present some preliminary results for $N=6$ in appendix \ref{dual-S5}.
\section{Conclusions}
Our results for $O(N)$ sigma models has to be extended for other integrable sigma models including supersymmetric ones.  An immediate generalization is related to the superalgebra $\mathfrak{sl}(n|n)$. In this case, the arguments of our paper can be repeated  with mild modification. We plan to return to this question in a future publication. As another example of the duality, it would be interesting to study the integrable deformations of strings on various supergravity backgrounds and find their dual description.
\section*{Acknowledgments}
The results of section \ref{CFT} and \ref{ricci} as well as of the appendix \ref{dual-S5} are obtained by A.L. The results of sections \ref{dual-L} and  of appendices \ref{IM} and \ref{group-element} are obtained by L.S. The introduction was written in close collaboration between the participants of this publication.

A.L. acknowledges discussions with Vladimir Fateev, Borya Feigin, Misha Bershtein and Serezha Lukyanov, his work is  supported by Laboratory of Mirror Symmetry NRU HSE, RF Government grant, ag. N 14.641.31.0001. The work of L.S. is supported by the Russian Science Foundation under grant 18-12-00439 and performed in Landau Institute for Theoretical Physics.
\Appendix
\section{Integrals of Motion}\label{IM}
Here we will study the set of commutativity equations \eqref{free-IOM} in more details.
Namely, let $\varphi(z)$ be the $(N-1)$ component bosonic chiral field $\varphi(z)=(\varphi_{1}(z),\dots,\varphi_{N-1}(z))$ normalized according to the operator product expansion
\begin{equation}
  \varphi_{i}(z)\varphi_{j}(w)=-\delta_{ij}\log(z-w)+\dots
\end{equation}
We note that \eqref{free-IOM} is trivially satisfied by the first Integral of Motion
\begin{equation}
    \mathbf{I}_{1}^{\textrm{free}}=\frac{1}{2\pi}\int_{0}^{2\pi}(\partial\varphi(z),\partial\varphi(z))dz.
\end{equation}
Nontrivial equations appear when we consider  Integral of Motion of higher spin. We assume that $\mathbf{I}_{2}^{\textrm{free}}=0$ and
\begin{equation}
   \mathbf{I}_{3}^{\textrm{free}}=\frac{1}{2\pi}\int_{0}^{2\pi}\mathbf{G}_{4}(z)dz\neq0.
\end{equation}
The density $\mathbf{G}_{4}(z)$ is defined up to a total derivative. It can be represented as
\begin{equation}\label{W4-generic}
 \mathbf{G}_{4}=A_{ijkl}\partial\varphi_{i}\partial\varphi_{j}\partial\varphi_{k}\partial\varphi_{l}+B_{ijk}\partial\varphi_{i}\partial\varphi_{j}\partial^{2}\varphi_{k}+
 C_{ij}\partial^{2}\varphi_{i}\partial^{2}\varphi_{j},
\end{equation}
with unknown tensors $A_{ijkl}$, $B_{ijk}$ and $C_{ij}$ with prescribed symmetry. The OPE of the field \eqref{W4-generic} with the exponential field $V_{\boldsymbol{\alpha}}(w)=e^{(\boldsymbol{\alpha},\varphi(w))}$, $\boldsymbol{\alpha}=(\alpha_{1},\dots,\alpha_{N-1})$ has the form
\begin{multline}\label{W4-OPE}
   \mathbf{G}_{4}(z)V_{\boldsymbol{\alpha}}(w)=\frac{wV_{\boldsymbol{\alpha}}(w)}{(z-w)^{4}}+\frac{\nu_{i}\partial\varphi_{i}(z)}{(z-w)^{3}}V_{\boldsymbol{\alpha}}(w)+
   \frac{\lambda_{ij}\partial\varphi_{i}(z)\partial\varphi_{j}(z)+\kappa_{i}\partial^{2}\varphi_{i}(z)}{(z-w)^{2}}V_{\boldsymbol{\alpha}}(w)+\\+
   \frac{\sigma_{ijk}\partial\varphi_{i}(z)\partial\varphi_{j}(z)\partial\varphi_{k}(z)+\rho_{ij}\partial\varphi_{i}(z)\partial^{2}\varphi_{j}(z)}{(z-w)}V_{\boldsymbol{\alpha}}(w)+\dots
\end{multline}
where the fields in the r.h.s. are Wick ordered. The tensors in the r.h.s. of \eqref{W4-OPE} are
\begin{equation}
\begin{gathered}
    \omega=A_{ijkl}\alpha_{i}\alpha_{j}\alpha_{k}\alpha_{l}+B_{ijk}\alpha_{i}\alpha_{j}\alpha_{k}+C_{ij}\alpha_{i}\alpha_{j},\quad
  \nu_{i}=-4A_{ijkl}\alpha_{j}\alpha_{k}\alpha_{l}-2B_{ijk}\alpha_{j}\alpha_{k},\\
  \lambda_{ij}=6A_{ijkl}\alpha_{k}\alpha_{l}+B_{ijk}\alpha_{k},\quad \kappa_{i}=B_{jki}\alpha_{j}\alpha_{k}+2C_{ij}\alpha_{j},\quad
  \sigma_{ijk}=-4A_{ijkl}\alpha_{l},\quad \rho_{ij}=-2B_{ikj}\alpha_{k}
\end{gathered}
\end{equation}
Using this OPE we can compute the integral
\begin{equation}\label{W4-OPE-integrated}
   \frac{1}{2\pi i}\oint_{\mathcal{C}_{z}}\mathbf{G}_{4}(z)V_{\boldsymbol{\alpha}}(w)dw=-\bigl(\tilde{\sigma}_{ijk}\partial\varphi_{i}(z)\partial\varphi_{j}(z)\partial\varphi_{k}(z)+
   \tilde{\rho}_{ij}\partial\varphi_{i}(z)\partial^{2}\varphi_{j}(z)\bigr)V_{\boldsymbol{\alpha}}(z)-\frac{\omega}{6}\partial^{3}V_{\boldsymbol{\alpha}}(z),
\end{equation}
where
\begin{equation*}
 \tilde{\sigma}_{ijk}=\sigma_{ijk}+\frac{1}{6}(\alpha_{i}\alpha_{j}\nu_{k}+\alpha_{i}\alpha_{k}\nu_{j}+\alpha_{j}\alpha_{k}\nu_{i})
 -\frac{1}{3}(\lambda_{ij}\alpha_{k}+\lambda_{ik}\alpha_{j}+\lambda_{jk}\alpha_{i}),\quad
 \tilde{\rho}_{ij}=\rho_{ij}+\frac{1}{2}\nu_{i}\alpha_{j}-\alpha_{i}\kappa_{j}.
\end{equation*}
We note that the r.h.s. of \eqref{W4-OPE-integrated} has to be total derivative. This condition leads to
\begin{equation}\label{sigma-rho-equations}
   \tilde{\rho}_{ij}=\tilde{\rho}_{ji},\qquad
   \tilde{\sigma}_{ijk}=\frac{1}{6}\left(\tilde{\rho}_{ij}\alpha_{k}+\tilde{\rho}_{ik}\alpha_{j}+\tilde{\rho}_{jk}\alpha_{i}\right).
\end{equation}
According to \eqref{free-IOM} these equations should be valid for all $\boldsymbol{\alpha}_{r}$ for $r=1,\dots,N$. For two sets of vectors \eqref{vectors} and \eqref{vectors-2} equations \eqref{sigma-rho-equations} can be solved. Their solution provide explicit form of the density $G_{4}(z)$. Explicitly (here $a=-i\sqrt{1+b^{2}}$ and all the monomial densities are Wick ordered)
\paragraph{Case $N=2n+1$:}
\begin{multline}\label{W4}
   \mathbf{G}_{4}(z)=\left((\partial\Phi,\partial\Phi)+(\partial\phi,\partial\phi)\right)^{2}+\frac{2n-1}{3}\sum_{k=1}^{n}\left(\frac{1}{b^{2}}(\partial\Phi_{k})^{4}+\frac{1}{a^{2}}(\partial\phi_{k})^{4}\right)+\\+
   2(2n-1)\sum_{k=1}^{n}\bigl((\partial\Phi_{k})^{2}+(\partial\phi_{k})^{2}\bigr)\biggl(\frac{1}{a}\sum_{j>k}\partial^{2}\Phi_{j}+\frac{1}{b}\sum_{j<k}\partial^{2}\phi_{j}-
    \frac{2}{2n-1}\sum_{j=1}^{n}(j-1)\Bigl(\frac{1}{a}\partial^{2}\Phi_{j}+\frac{1}{b}\partial^{2}\phi_{n-j+1}\Bigr)\biggr)+\\+
    \biggl(\frac{4(n+1)}{3}+\frac{2n-1}{3}\Bigl(\frac{1}{b^{2}}+\frac{2}{a^{2}}\Bigr)\biggr)(\partial^{2}\Phi,\partial^{2}\Phi)+
    \biggl(\frac{4(n+1)}{3}+\frac{2n-1}{3}\Bigl(\frac{2}{b^{2}}+\frac{1}{a^{2}}\Bigr)\biggr)(\partial^{2}\phi,\partial^{2}\phi)+\\+
    2\sum_{i\leq j}(i-1)(2(j-n)-1)\bigr(2-\delta_{ij}\bigr)\Bigl(\frac{1}{a^{2}}\partial^{2}\Phi_{i}\partial^{2}\Phi_{j}+\frac{1}{b^{2}}\partial^{2}\phi_{n-i+1}\partial^{2}
    \phi_{n-j+1}\Bigr)+\\+
    \frac{2}{a b}\Bigl(4\sum_{i,j}(i-1)(n-j)\partial^{2}\Phi_{i}\partial^{2}\phi_{j}-
    (2n-1)\sum_{i>j}(2(i-j)-1)\partial^{2}\Phi_{i}\partial^{2}\phi_{j}\Bigr).
\end{multline}
\paragraph{Case $N=2n+2$:}
\begin{multline}\label{W4-2}
   \mathbf{G}_{4}(z)=\left((\partial\Phi,\partial\Phi)+(\partial\phi,\partial\phi)\right)^{2}+\frac{2n}{3}\left(\sum_{k=1}^{n}\frac{1}{b^{2}}(\partial\Phi_{k})^{4}+\sum_{k=1}^{n+1}\frac{1}{a^{2}}(\partial\phi_{k})^{4}\right)+\\+
   4n\sum_{k=1}^{n}\bigl((\partial\Phi_{k})^{2}+(\partial\phi_{k})^{2}\bigr)\biggl(\frac{1}{a}\sum_{j>k}\partial^{2}\Phi_{j}+\frac{1}{b}\sum_{j<k}\partial^{2}\phi_{j}-
    \frac{1}{n}\Bigl(\frac{1}{a}\sum_{j=1}^{n}(j-1)\partial^{2}\Phi_{j}+\frac{1}{b}\sum_{j=1}^{n+1}(j-\frac{1}{2})\partial^{2}\phi_{n-j+2}\Bigr)\biggr)+\\+
    \biggl(\frac{4(2n+3)}{6}+\frac{2n}{3}\Bigl(\frac{1}{b^{2}}+\frac{2}{a^{2}}\Bigr)\biggr)(\partial^{2}\Phi,\partial^{2}\Phi)+
    \biggl(\frac{4(2n+3)}{6}+\frac{2n}{3}\Bigl(\frac{2}{b^{2}}+\frac{1}{a^{2}}\Bigr)\biggr)(\partial^{2}\phi,\partial^{2}\phi)+\\+
    2\sum_{i\leq j}\Bigl(\frac{1}{a^{2}}(i-1)(2(j-n)-2)(2-\delta_{ij})\partial^{2}\Phi_{i}\partial^{2}\Phi_{j}+
    \frac{1}{b^{2}}\bigl(i-\frac{1}{2}\bigr)(2(j-n)-1)(2-\delta_{ij})\partial^{2}\phi_{n-i+2}\partial^{2}
    \phi_{n-j+2}\Bigr)+\\+
    \frac{2}{a b}\Bigl(4\sum_{i,j}(i-1)\left(\frac{2n+1}{2}-j\right)\partial^{2}\Phi_{i}\partial^{2}\phi_{j}-
    n\sum_{i>j}(2(i-j)-1)\partial^{2}\Phi_{i}\partial^{2}\phi_{j}\Bigr).
\end{multline}

From \eqref{W4} and \eqref{W4-2} we note that
\begin{equation}
   \mathbf{G}_{4}=\left((\partial\Phi,\partial\Phi)+(\partial\phi,\partial\phi)\right)^{2}+\frac{2(N+2)}{3}\left((\partial^{2}\Phi,\partial^{2}\Phi)+(\partial^{2}\phi,\partial^{2}\phi)\right)
   +O\left(\frac{1}{b^{2}}\right)\quad\text{at}\quad b\rightarrow\infty
\end{equation}
We see that in the leading order at $b\rightarrow\infty$  the density $G_{4}$ enjoys the $O(N-1)$ symmetry  and coincides with the density studied in \cite{Lukyanov:2003rt}.
\section{Parametrization of the group elements}\label{group-element}
The action of the deformed coset sigma model has the form
\begin{equation}\label{Coset-action-deformed-app}
   \mathcal{S}=\frac{1}{2}\int\textrm{Tr}\left(
   \left(\mathbf{g}\partial_{+}\mathbf{g}^{-1}\right)^{(\textrm{c})}\,\frac{1}{1-i\kappa\mathcal{R}_{\mathbf{g}}\circ\mathrm{P}_{\textrm{c}}}\,
   \left(\mathbf{g}\partial_{-}\mathbf{g}^{-1}\right)^{(\textrm{c})}\right) d^{2}x,
\end{equation}
We choose the basis in the Lie algebra $SO(N)$ to be
\begin{equation}
   (T_{ab})_{ij}\overset{\text{def}}{=}\delta_{ai}\delta_{bj}-\delta_{bi}\delta_{aj},
\end{equation}
and $SO(N-1)$ subalgebra is chosen to be $T_{ab}$ with $a,b\ne 1$. Now we have to choose some parametrization of the group element\footnote{Here we follow \cite{Hoare:2015gda}.}.
\subsection*{N=4}
In this case, we choose the following parametrization of $SO(4)/SO(3)$
\begin{equation}
\mathbf{g}^{-1}=\exp\left(\phi_1 \,T_{34}\right)\exp\left(\phi\, T_{12}\right)\exp\left(\arcsin \zeta\, T_{13}\right)
\end{equation}
The metric reads
\begin{equation}\label{Sausage-metric-app}
   ds^{2}=\frac{1}{2}\left(\frac{d\zeta^{2}}{(1-\zeta^{2})(1-\kappa^{2}\zeta^{2})}+\frac{(1-\zeta^{2})d\phi^{2}}{(1-\kappa^{2}\zeta^{2})}+\zeta^2 d\phi_1^2\right).
\end{equation}
$N=3$ cases is the same with the coordinate $\phi_1$ set to zero.
\subsection*{N=6}
In this case we choose the following paramterization of $SO(6)/SO(5)$
\begin{equation}
\mathbf{g}^{-1}=\exp\left(\phi_2 \,T_{56}\right)\exp\left(\phi_1 \,T_{34}\right)\exp\left(\theta \,T_{35}\right)\exp\left(\phi\, T_{12}\right)\exp\left(\arcsin \zeta\, T_{13}\right)
\end{equation}
The metric reads
\begin{align}
  ds^{2}=\frac{1}{2}\left(\frac{d\zeta^{2}}{(1-\zeta^{2})(1-\kappa^{2}\zeta^{2})}+\frac{\zeta^2}{1-\kappa^2 \zeta^4 \sin^2 \theta}+\frac{(1-\zeta^{2})d\phi^{2}}{(1-\kappa^{2}\zeta^{2})}+\frac{\zeta^2 \cos^2\theta}{1-\kappa^2 \zeta^4 \sin^2\theta}+\zeta^2 \sin^2\theta d\phi_2^2\right).
\end{align}
The $N=5$ case is achieved by setting $\phi_2$ to zero.

Besides the pure metrics term the integrability requires to add the $B$-field. The additional term in the action is
\begin{equation}
   \mathcal{S_B}=\frac{1}{2}\int\textrm{Tr}\left(\epsilon_{ab}
   \left(\mathbf{g}\partial_{a}\mathbf{g}^{-1}\right)^{(\textrm{c})}\,\frac{1}{1-i\kappa\mathcal{R}_{\mathbf{g}}\circ\mathrm{P}_{\textrm{c}}}\,
   \left(\mathbf{g}\partial_{b}\mathbf{g}^{-1}\right)^{(\textrm{c})}\right) d^{2}x,
\end{equation}
and direct computation shows the corresponding $B$-field is
\begin{align}\label{B-field-app}
B=\frac{i\kappa \zeta }{1-\zeta^2\kappa^2}d\zeta\wedge d\phi+\frac{i\kappa \zeta^4\sin\theta\cos\theta}{1-\kappa^2\zeta^4  \sin^2 \theta}d\theta\wedge d \phi_1
\end{align}
One can easily see that the first term in \eqref{B-field-app} is a pure gauge and can be discarded.
\section{Dual metric for the deformed $O(6)$ sigma model}\label{dual-S5}
Consider sigma model with $5$-dimensional target space which is dual with respect to $b^{2}\rightarrow-1-b^{2}$ to the theory studied in section \ref{ricci}. It corresponds to the Dynkin graph
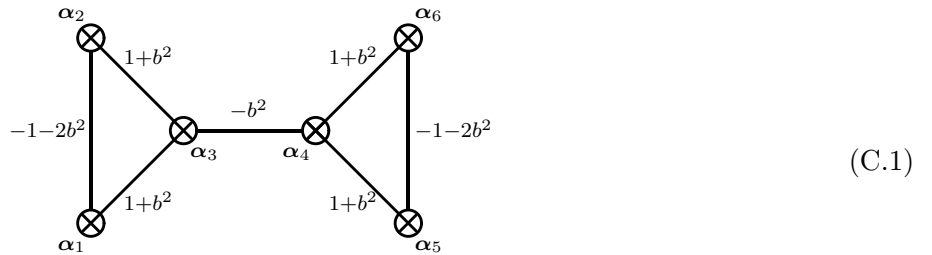
\begin{equation}\label{D3-Dynkin}
\begin{picture}(300,60)(160,110)
    \Thicklines
    \unitlength 5pt
    \put(48,32){\circle{2}}
    \put(48,18){\circle{2}}
    \put(54.4,24,4){\line(-1,-1){7}}
    \put(54.4,25,6){\line(-1,1){7}}
    \put(47.4,31.4){\line(1,1){1.2}}
    \put(47.4,18.6){\line(1,-1){1.2}}
    \put(48,19){\line(0,1){12}}
    \put(55,25){\circle{2}}
    \put(54.4,24,4){\line(1,1){1.2}}
    \put(54.4,25,6){\line(1,-1){1.2}}
    \put(56,25){\line(1,0){8}}
    \put(65,25){\circle{2}}
    \put(64.4,24,4){\line(1,1){1.2}}
    \put(64.4,25,6){\line(1,-1){1.2}}
    \put(64.4,24,4){\line(1,1){8.2}}
    \put(64.4,25,6){\line(1,-1){8.2}}
    \put(72,32){\circle{2}}
    \put(72,18){\circle{2}}
    \put(71.4,32.6){\line(1,-1){1.2}}
    \put(71.4,17.4){\line(1,1){1.2}}
    \put(72,19){\line(0,1){12}}
    \put(50.5,19){$\scriptstyle{1+b^{2}}$}
    \put(50.5,30){$\scriptstyle{1+b^{2}}$}
    \put(45.5,16){$\scriptstyle{\boldsymbol{\alpha}_{1}}$}
    \put(45.5,33.5){$\scriptstyle{\boldsymbol{\alpha}_{2}}$}
    \put(55.5,23){$\scriptstyle{\boldsymbol{\alpha}_{3}}$}
    \put(62.5,23){$\scriptstyle{\boldsymbol{\alpha}_{4}}$}
    \put(72.5,16){$\scriptstyle{\boldsymbol{\alpha}_{5}}$}
    \put(72.5,33.5){$\scriptstyle{\boldsymbol{\alpha}_{6}}$}
    \put(66,19){$\scriptstyle{1+b^{2}}$}
    \put(66,30){$\scriptstyle{1+b^{2}}$}
    \put(41.8,24.5){$\scriptstyle{-1-2b^{2}}$}
    \put(72.5,24.5){$\scriptstyle{-1-2b^{2}}$}
    \put(58.5,26){$\scriptstyle{-b^{2}}$}
  \end{picture}
  \vspace*{1cm}
\end{equation}
The vectors $\boldsymbol{\alpha}_{k}$ can be parametrized as follows
\begin{equation}
\begin{aligned}
    &\boldsymbol{\alpha}_{1}=i\beta e_{1}+b E_{1},\quad&&\boldsymbol{\alpha}_{2}=i\beta e_{1}-b E_{1},\quad&&\boldsymbol{\alpha}_{3}=-i\beta e_{1}+b E_{2},\\
    &\boldsymbol{\alpha}_{4}=-i\beta e_{2}-b E_{2},\quad&&\boldsymbol{\alpha}_{5}=i\beta e_{2}+b E_{3},\quad&&\boldsymbol{\alpha}_{6}=i\beta e_{2}-b E_{3}.
\end{aligned}
\end{equation}

We look for the solution to  \eqref{RG-equation-3} in the form
\begin{equation}\label{anzatz-4}
  G_{\mu\nu}=\delta_{\mu\nu}+e^{\alpha t}\left(A_{\mu\nu}e^{x_{2}+x_{1}}+A^{+}_{\mu\nu}e^{x_{2}-x_{1}}+B_{\mu\nu}e^{-x_{2}+x_{3}}+B^{+}_{\mu\nu}e^{-x_{2}-x_{3}}\right)+\dots,\quad
  \Psi=(\rho,x)+\dots,
\end{equation}
where
\begin{equation*}
  A_{\mu\nu}=\begin{pmatrix}
  1&0&0&i&0\\
  0&0&0&0&0\\
  0&0&0&0&0\\
  i&0&0&-1&0\\
  0&0&0&0&0
  \end{pmatrix},\quad
  B_{\mu\nu}=\begin{pmatrix}
  0&0&0&0&0\\
  0&0&0&0&0\\
  0&0&1&0&i\\
  0&0&0&0&0\\
  0&0&i&0&-1
  \end{pmatrix}.
\end{equation*}
It corresponds to the perturbation by the operators
\begin{equation}
\begin{gathered}
  \int(\boldsymbol{\alpha}_{1},\partial\varphi)(\boldsymbol{\alpha}_{1},\bar{\partial}\varphi)e^{(\boldsymbol{\beta}_{13},\varphi)}d^{2}z,\quad
  \int(\boldsymbol{\alpha}_{2},\partial\varphi)(\boldsymbol{\alpha}_{2},\bar{\partial}\varphi)e^{(\boldsymbol{\beta}_{23},\varphi)}d^{2}z,\\
  \int(\boldsymbol{\alpha}_{5},\partial\varphi)(\boldsymbol{\alpha}_{5},\bar{\partial}\varphi)e^{(\boldsymbol{\beta}_{45},\varphi)}d^{2}z,\quad
  \int(\boldsymbol{\alpha}_{6},\partial\varphi)(\boldsymbol{\alpha}_{6},\bar{\partial}\varphi)e^{(\boldsymbol{\beta}_{46},\varphi)}d^{2}z.
\end{gathered}
\end{equation}
At the leading order we obtain
\begin{equation}
   \alpha=1,\qquad \rho=(0,0,0,-\frac{i}{2},-\frac{i}{2})
\end{equation}
We also assume that the determinant of the metric $G_{\mu\nu}$ is an integral of motion. Then one can check that the solution stays within the following  anzatz
\begin{equation*}
   \begin{pmatrix}
  \coth F+\frac{\cosh x_{1}}{\sinh F}&0&0&i\frac{\sinh x_{1}}{\sinh F}&0\\
  0&\phi&0&0&0\\
  0&0&\coth \bar{F}+\frac{\cosh x_{3}}{\sinh\bar{F}}&0&i\frac{\sinh x_{3}}{\sinh\bar{F}}\\
  i\frac{\sinh x_{1}}{\sinh F}&0&0&\coth F-\frac{\cosh x_{1}}{\sinh F}&0\\
  0&0&i\frac{\sinh x_{3}}{\sinh\bar{F}}&0&\coth \bar{F}-\frac{\cosh x_{3}}{\sinh\bar{F}}\\
  \end{pmatrix},
\end{equation*}
and $\Psi=-\frac{i}{2}(x_{4}+x_{5})+\Phi$. The functions $F,\bar{F}$ and $\Phi$ depend on $x_{2}$ and $t$ only, while $\phi$ depends only on $t$. The Ricci flow equations  \eqref{RG-equation-3} in this case are reduced to the system of PDE's
\begin{equation}\label{S5-dual-eqs}
\begin{gathered}
       \Phi''=\frac{1}{4}\left(\frac{F'^{2}}{\sinh^{2}F}+\frac{\bar{F}'^{2}}{\sinh^{2}\bar{F}}-2\dot{\phi}\right),\\
        F''=2F'\Phi'+F'^{2}\coth F+\phi(2\dot{F}+1),\quad
        \bar{F}''=2\bar{F}'\Phi'+\bar{F}'^{2}\coth\bar{F}+\phi(2\dot{\bar{F}}+1).
\end{gathered}
\end{equation}
In our case the special symmetry holds
\begin{equation}
  \bar{F}(x_{2},t)=F(-x_{2},t),\qquad \Phi(x_{2},t)=\Phi(-x_{2},t),
\end{equation}
so the last equation in \eqref{S5-dual-eqs} can be dropped. Analyzing first terms in the expansion  at $t\rightarrow-\infty$, one can find  that
\begin{equation}\label{Dual-S5-answer}
   \phi=-\left(\frac{2K(m)}{\pi}\right)^{2}\coth t,\quad
   F=-t+\log\bigl(\textrm{cn}(iz|m)+i\,\textrm{sn}(iz|m)\bigr),
\end{equation}
where
\begin{equation*}
    z=\frac{2K(m)}{\pi}x_{2},\qquad m=-\frac{1}{\sinh^{2}t},
\end{equation*}
and $K(z)$, $\textrm{cn}(z|m)$, $\textrm{sn}(z|m)$ are the standard elliptic functions.
\bibliographystyle{MyStyle}
\bibliography{MyBib}

\end{document}